\documentclass[letterpaper,11pt]{article}
\usepackage[T1]{fontenc}
\usepackage[margin=3cm]{geometry}
\usepackage[affil-it]{authblk}

\usepackage{amssymb,amsmath,amsthm,amsfonts}

\usepackage{listings}

\usepackage{algorithm}
\usepackage{algorithmicx}
\usepackage[noend]{algpseudocode}
\usepackage[font=small,labelfont=bf]{caption,subcaption}

\usepackage[table]{xcolor}
\usepackage{graphicx}

\usepackage{booktabs}
\usepackage{tabularx}
\usepackage{hyperref}

\usepackage[noabbrev,capitalize]{cleveref}
\usepackage{url}
\usepackage{xurl}

\definecolor{yellow1}{rgb}{1,0.55,0} 
\definecolor{green1}{rgb}{0,0.4,0} 
\definecolor{blue1}{rgb}{0,0.1880,1}

\usepackage[backend=biber, sorting=none, style=numeric, url=false, maxbibnames=99, doi=false, url=false, isbn=false]{biblatex}
\usepackage{multirow}
\usepackage{scalerel} 
\usepackage{csquotes}

\usepackage{verbatim}
\makeatletter
\newcommand{\verbatimfont}[1]{\def\verbatim@font{#1}}%
\makeatother
\verbatimfont{\ttfamily\small}

\usepackage{sty/macros}

\usepackage[most]{tcolorbox}
\newtcolorbox{algbackbox}{width=\textwidth,top=0pt,left=0pt,right=0pt,bottom=0pt,toptitle=0pt,bottomtitle=0pt, colback=white,colframe=white}
\tcbset{colback=gray!10!white,colframe=black,colbacktitle=gray!40!white,coltitle=black,fonttitle=\large\bfseries, before skip=6mm}

\makeatletter
\newcommand{\settitle}{\@maketitle}
\makeatother

\title{\Huge Nonlinear System Identification\\\Large Learning while respecting physical models\\ using a sequential Monte Carlo method}

\author[1]{Anna Wigren\thanks{\url{anna.wigren@it.uu.se}}}
\author[2]{Johan W{\aa}gberg\thanks{\url{johan.wagberg@it.uu.se}}}
\author[3]{Fredrik Lindsten\thanks{\url{fredrik.lindsten@liu.se}}}
\author[4]{Adrian Wills\thanks{\url{adrian.wills@newcastle.edu.au}}}
\author[5]{Thomas B. Sch{\"o}n\thanks{\url{thomas.schon@it.uu.se}}}
\affil[1,2,5]{Department of Information Technology, Uppsala University}
\affil[3]{Department of Computer and Information Science, Link{\"op}ing University}
\affil[4]{School of Engineering, University of Newcastle}
\date{}

\begin{document}

\thispagestyle{empty}
\addtocounter{page}{-1}

\begin{center}
	{\Huge Nonlinear System Identification\\\Large Learning while respecting physical models\\ using a sequential Monte Carlo method }
\end{center}

\vspace{-1mm}

\begin{center}
	Anna Wigren, Johan Wågberg, Fredrik Lindsten, Adrian G. Wills, Thomas B. Schön
\end{center}

\vspace{1cm}

\noindent \textbf{Please cite this version:}

\noindent Anna Wigren, Johan Wågberg, Fredrik Lindsten, Adrian G. Wills, Thomas B. Schön. “Nonlinear System Identification: Learning While Respecting Physical Models Using a Sequential Monte Carlo Method.” In: IEEE Control Systems Magazine 42.1 (2022). © 2022 IEEE, pp. 75–102

\begin{center}
	\begin{minipage}{.9\textwidth}
		\begin{lstlisting}[breaklines,basicstyle=\small\ttfamily]
@article{Wigren2022,
author={Wigren, Anna and W{\aa}gberg, Johan and Lindsten, Fredrik and Wills, Adrian G. and Sch{\"o}n, Thomas B.},
journal={IEEE Control Systems Magazine}, 
title={Nonlinear System Identification: Learning While Respecting Physical Models Using a Sequential Monte Carlo Method}, 
year={2022},
volume={42},
number={1},
pages={75-102},
doi={https://doi.org/10.1109/MCS.2021.3122269},
}		
		\end{lstlisting}
	\end{minipage}
\end{center}

\vspace{1cm}

\noindent \textbf{A note on the structure of the article:}

\noindent The published version of this article consists of a main text that provides the essential content and multiple sidebars with additional information, either in the form of examples or a background with further technical details. The same structure has been adopted in this version of the article. Sidebars are indicated by grey boxes and are referenced from the main text using double quotation marks, i.e. ``Background: Markov chain Monte Carlo'' refers to the sidebar on Markov chain Monte Carlo. The sidebars are placed at the end of the section where they are first referenced.

\maketitle

\begin{abstract}
	Identification of nonlinear systems is a challenging problem.
Physical knowledge of the system can be used in the identification process to significantly improve the predictive performance by restricting the space of possible mappings from the input to the output.
Typically, the physical models contain unknown parameters that must be learned from data.
Classical methods often restrict the possible models or have to resort to approximations of the model that introduce biases.
Sequential Monte Carlo methods enable learning without introducing any bias for a more general class of models.
In addition, they can also be used to approximate a posterior distribution of the model parameters in a Bayesian setting. 
This article provides a general introduction to sequential Monte Carlo and shows how it naturally fits in system identification by giving examples of specific algorithms.
The methods are illustrated on two systems: a system with two cascaded water tanks with possible overflow in both tanks and a compartmental model for the spreading of a disease.
\end{abstract}

\section{Introduction}
The modern world contains an immense number of different and interacting systems, from the evolution of weather systems to variations in the stock market, autonomous vehicles interacting with their environment and the spread of diseases.
For society to function, it is essential to understand the behavior of the world, so that informed decisions can be made that are based on likely future outcomes. For instance, consider the spread of a new disease like the coronavirus.
It is of great importance to be able to \textit{predict} the number of people that will be infected at different points in time to ensure that appropriate healthcare facilities are available.
It is also of interest to be able to make \textit{decisions} based on accurate information to best attenuate the spread of disease.
Moreover, \textit{understanding} specific attributes of the disease, such as the incubation time, the number of unreported cases, and how \textit{certain} we are about this knowledge are also crucial.

These types of applications are examples of so-called \emph{dynamic} systems, which are the focus of this article.
Dynamic systems have the property that the future system response depends on the past system response~\cite{Ljung:1999}.
Capturing these types of dynamic phenomena can be achieved using mathematical models, which offer a concrete mechanism for making predictions and supporting decisions.
The extreme flexibility and versatility of mathematics affords modeling of highly disparate dynamic behavior. However, it also creates a challenge, since it is not always obvious how to choose an appropriate model.
This diversity is perhaps best illustrated by contrasting examples. 

Consider the modeling of rigid-body vehicle dynamics, such as the motion of a car or a plane.
In this case, it is possible to exploit \emph{prior knowledge} of the system and adopt a classical mechanics approach to derive Newton-Euler equations of motion for each application~\cite{analmek}.
The mathematical model structure is largely determined by knowledge of the physical system, and the model will depend on certain parameters such as mass and inertia terms, and damping and friction coefficients.
In many cases, these parameter values can be difficult to obtain based on first principles approaches alone.
It is important to also note that some parameters may have feasible ranges, such as mass terms being nonnegative, which is also a form of prior knowledge.

Contrasting this type of model, it is also possible to employ highly flexible and general model structures to describe dynamic systems, such as deep neural networks (DNNs) or Gaussian processes (GPs)~\cite{GoodfellowBC:2016,ljung2020deep,TurnerDR:2009}.
The flexibility of the DNN model class stems from the general construction of the model, which involves potentially many layers of interacting nonlinear functions.
Importantly, these interactions are allowed to adapt for each new application, since they rely on coefficients/parameters that are free to change values.
In the case of GP, the model structure is also highly flexible, nonparametric and adapted based on available data.
For either of these flexible model classes, it is more challenging to impose prior system knowledge. However, some progress is being made along these lines~\cite{JidlingWWS:2017,HendriksJWS:2020}.

Irrespective of the type of model, there are unknown quantities that must be determined, which are often inferred from observations from the system that is being modeled.
There are many different approaches for extracting or estimating these unknown values from observed system data~\cite{Rubin:1984,Fisher:1912,Fisher:1922,Bayes:1763,godambe1991estimating}.
Among the many possibilities, this article concentrates on two commonly used and complimentary approaches.
In particular, the presented inference methods are grouped according to two main attributes: 1) the assumptions made about how to model unknown parameters, and, 2) what should be estimated in addition to the parameters.  

More specifically, if the model parameters are assumed to be deterministic variables, then this results in a frequentist inference perspective, where the so-called \emph{maximum likelihood} (ML) approach has proven to be highly successful in providing accurate point estimates of the parameters~\cite{Fisher:1912,Fisher:1922}.
Alternately, if uncertainty about the parameter values is incorporated by treating them as random variables, then this results in the so-called Bayesian perspective, where the \textit{posterior distribution} of the parameters is the object of interest~\cite{Bayes:1763}.
An attractive attribute of the Bayesian approach is that it provides quantification of uncertainty, which is essential when making decisions based on the associated models.
Otherwise, decisions may be executed based on misplaced confidence.
It is also worth mentioning that there is a connection between these two approaches by considering so-called \emph{maximum a posteriori} methods~\cite{Bassett:2019}. 

Regardless of adopting the frequentist or Bayesian perspective, it is rare that the estimates can be provided analytically.
This article provides computational tools for calculating these estimates in the remaining cases where analytical solutions are not available.
Towards computing them, it is essential to both the frequentist and Bayesian approaches that certain integrals can be evaluated.
While the details will be explained in subsequent sections, it suffices for now to mention that computing these integrals is generally intractable~\cite{Kantas:2015}. 

An overarching theme of this article is to approximate intractable integrals by employing carefully tailored Monte-Carlo integration techniques that result in tractable weighted sums.
More precisely, the sequential nature of the dynamic models lends itself to the so-called
sequential Monte Carlo (SMC) methods~\cite{Gordon:1993,Delmoral:1996}, which will be explored in much more detail as the article progresses.
Furthermore, these SMC methods are employed both within frequentist and Bayesian approaches, resulting in algorithms that are applicable to a wide range of modeling problems.
An attractive property of the SMC methods is that they also offer asymptotic convergence guarantees, which are not offered by other approximation methods in general~\cite{Kantas:2015}.

These SMC techniques are also highly suitable to the situation where \emph{prior knowledge} of the system is available, such as knowledge of the physical system, model structure, and possibly feasible ranges for unknown parameter values.
The main benefit of using SMC techniques is that they are applicable to general nonlinear systems, without modification of the prior knowledge or assumptions.
This allows the separation of modeling from the inference method, which provides the modeler more freedom in adding domain-specific prior knowledge.
In contrast, many alternative approaches require\textemdash either explicitly or implicitly\textemdash that the problem satisfies certain restrictive assumptions, such as Gaussian noise corruption.
The aim of the article is to present computational tools for estimating general nonlinear dynamic systems, while adhering to prior system knowledge without modification.

The article first presents the type of models considered and provides two examples that illustrate how physical insight about the system can be transformed to a mathematical model suited for statistical inference.
These two examples are used throughout to illustrate the various methods.
Following this, the identification problem is introduced, and the key expressions needed for learning are highlighted.
This leads to an introduction of SMC specifically targeted for offline system identification.
The remainder of the article presents identification algorithms where SMC plays an integral part.
Both optimization-based learning methods and probabilistic methods where posterior distributions are computed are considered and applied to the example models.
The article also gives a short introduction to probabilistic programming, a tool that can significantly reduce the complexity of trying out different models and inference methods.


\section{Modeling}
Mathematical modeling is applicable to a wide range of problems spanning many areas of science and engineering.
As such, it is important to restrict attention to the particular model class of interest to this article, namely, discrete-time state-space models for dynamic systems.
These types of models have a long and fruitful history in the fields of physics and engineering, originating in the phase-space ideas from physics \cite{nolte2010tangled}.
The essential idea is that the dynamic behavior of the model is determined by the current \emph{state} of the model, which is a vector belonging to a so-called state space.
It is important to mention that the states should be associated with the model, rather than the real-world phenomena.
The latter has no particular concern for states or any other modeling choices, including the model structure and associated parameters. 

It is essential to connect observations from the real-world phenomena to the state-space model, since this is the primary purpose of modeling, and so that the model can be adapted to best match observations.
These ideas are made more concrete in the subsequent section, which introduces the state-space model of interest in this article, and presents two concrete examples to illustrate this modeling approach.

\subsection{Probabilistic formulation of the state-space model}
To make the modeling ideas discussed above concrete, it is necessary to introduce some notation.
To that end, the model state is denoted $\vstate_{\tv}$, where the subscript ${\tv}$ indicates the current discrete time instant.
Observations from the system are denoted $\vobs_{\tv}$, and inputs to the system are denoted $\vinp_{\tv}$.
It is typical to express the connection between model and observations via the state-space equations
\begin{subequations}
	\begin{align}
		\vstate_{\tv} &= f_{\tv}(\vstate_{\tv-1},\vinp_{\tv},\vpnoise_{\tv},\vparam), \\
		\vobs_{\tv} &= g_{\tv}(\vstate_{\tv},\vinp_{\tv}, \vonoise_{\tv},\vparam).
	\end{align}	
\end{subequations}
In the above, the function $f_{\tv}$ explains how the state evolves over time, and $g_{\tv}$ relates the model state to the system observations.
The parameter vector $\vparam$ allows the functions to depend on some possibly unknown parameters, and $\vpnoise_{\tv}$ and $\vonoise_{\tv}$ are noise terms to account for uncertainty. As an example, the functional form of a linear-Gaussian state-space model is
\begin{subequations}
\label{eq:LGfunctional}
\begin{alignat}{2} 
	\vstate_{\tv} &= \mat{A} \vstate_{\tv-1} + \mat{B} \vinp_{\tv} + \vpnoise_{\tv},\quad & \vpnoise_{\tv}&\sim \N{\vpnoise_{\tv} \mid \0,\mat{Q}}, \label{eq:LG-transition}\\
	\vobs_{\tv} &= \mat{C} \vstate_{\tv} + \mat{D} \vinp_{\tv} + \vonoise_{\tv}, & \vonoise_{\tv}&\sim \N{\vonoise_{\tv} \mid \0,\mat{R}}, \label{eq:LG-observation}
\end{alignat}
\end{subequations}
where $\mat{A}$ is a transition matrix, $\mat{B}$ is an input matrix, $\mat{C}$ is an observation matrix, $\mat{D}$ is a feedforward matrix, and $\vpnoise_{\tv}$ and $\vonoise_{\tv}$ are independent and identically distributed Gaussian noise with zero mean and covariance matrices $\mat{Q}$ and $\mat{R}$, respectively.
The unknown parameters of this model are the transition matrix $\mat{A}$, the input matrix $\mat{B}$, the observation matrix $\mat{C}$, the feedforward matrix $\mat{D}$, and the covariance matrices $\mat{Q}$ and $\mat{R}$.
The notation $\N{\vec{z} \mid \vec{\mu}, \mat{\Sigma}}$ is used to denote a multivariate Gaussian distribution with mean $\vec{\mu}$ and covariance matrix $\mat{\Sigma}$ for the variable $\vec{z}$.

\begin{figure}[htb]
	\centering
	\includegraphics[width=0.5\textwidth]{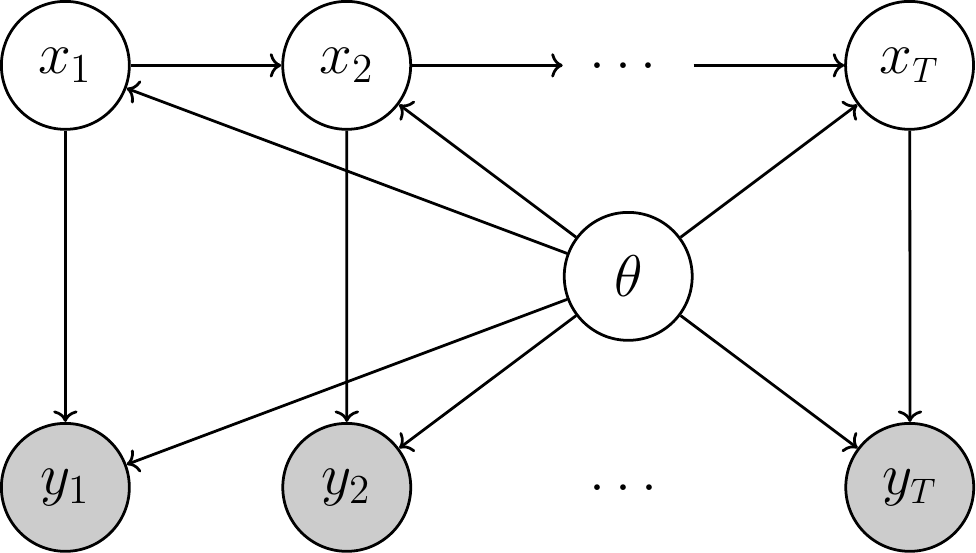}
	\caption{
		A probabilistic graphical model representation of the state-space model.
		In a probabilistic graphical model, each random variable is denoted by a circle.
		If the variable is observed, the circle is shaded in gray.
		Each arrow indicates a dependency. For example, the state at time $2$ depends on the state at time $1$, and all states $\vstate_{\range{1}{T}}$ and measurements $\vobs_{\range{1}{T}}$ depend on the parameters $\vparam$.
	}
	\label{fig:SSM}
\end{figure}

This article uses a more general, probabilistic, form of the state-space model, where the essential idea remains the same: The state holds the information required to determine the state evolution.
The main difference is the manner in which this is expressed.
For probabilistic state-space models, the time evolution and measurement relationships are captured via the \emph{conditional} probability distributions
\begin{subequations}
	\label{eq:state-space model}
	\begin{align}
		\vstate_{\tv} &\sim p(\vstate_{\tv} \mid \vstate_{\tv-1},\vinp_{\tv},\vparam), \label{eq:state evoltuion}\\
		\vobs_{\tv} &\sim p(\vobs_{\tv} \mid \vstate_{\tv},\vinp_{\tv},\vparam),\label{eq:measurement equation}
	\end{align}
\end{subequations}
with transition density $p(\vstate_{\tv} \mid \vstate_{\tv-1}, \vinp_{\tv}, \vparam)$ and observation density $p(\vobs_{\tv} \mid \vstate_{\tv}, \vinp_{\tv}, \vparam)$, parameterized by an unknown parameter $\vparam$. 
A probabilistic state-space model can, equivalently, be represented graphically as a probabilistic graphical model.
\Cref{fig:SSM} illustrates the graphical representation of \cref{eq:state-space model}.
The probabilistic modeling approach can be generalized to a rich class of systems that extends well beyond state-space models.
A brief discussion of this general approach is available in ``Background: Probabilistic models in general''.

Continuing the above linear-Gaussian example, the equivalent probabilistic form of the model \cref{eq:LGfunctional} is
\begin{subequations}
\label{eq:LGexample}
\begin{align} \label{eq:LGexample1}
    \vstate_{\tv}
		&\sim p(\vstate_{\tv} \mid \vstate_{\tv-1}, \vinp_{\tv}, \vparam)
		= \N{\vstate_{\tv} \mid \mat{A} \vstate_{\tv-1} + \mat{B} \vinp_{\tv}, \mat{Q}},  \\
	\vobs_{\tv}
		&\sim p(\vobs_{\tv} \mid \vstate_{\tv}, \vinp_{\tv}, \vparam)
		= \N{\vobs_{\tv} \mid \mat{C} \vstate_{\tv} + \mat{D} \vinp_{\tv}, \mat{R}}. \label{eq:LGexamplee}
\end{align}
\end{subequations}
In addition to this model, the initial state $\vstate_{1}\sim \N{\vstate_{1} \mid \vec{\mu}_{1}, \mat{\Sigma}_1}$, where $\vec{\mu}_1$ is the mean and $\mat{\Sigma}_{1}$ is a covariance matrix, must also be specified.
Two examples of the functional as well as the probabilistic form for nonlinear state-space models are available in ``Example: Cascaded water tanks'' and ``Example: Dengue fever''.
In both cases, the model structure is adapted to the application at hand by incorporating ``physical insight'' about the application in the governing equations.
This is useful, not only for obtaining a model that respects the physical properties of the system, but also for enabling more efficient learning of model parameters, compared to using generic black-box models.

When an appropriate model structure has been determined, which may be an iterative process, then the focus can be placed on learning (or estimating or identifying) the unknown parameters $\vparam$ based on observed data.
This is the problem treated in the remainder of the article.

\begin{tcolorbox}[title=Background: Probabilistic models in general, breakable, enhanced, before upper={\parindent15pt\noindent}] 
	A probabilistic model describing the relation between two random variables $A \in \mathcal{A}$ and $B \in \mathcal{B}$ can be specified through the joint distribution of these variables $p_{A,B}(a,b)$.
The subscripts indicate which random variables that the distribution describes.
Apart from in this section, the subscripts are left out in the interest of space and notational simplicity.
The variable $A$ can, for instance, be observations from a system of interest, and $B$ can be parameters and/or state variables in the probabilistic model.

When working with probabilistic models, there are two basic operations that are used extensively to form other distributions from the joint distribution.
The first is the \emph{marginalized} distribution of one of the random variables, which is obtained from the joint distribution by simply integrating out all other variables.
Hence, for the variable $B$, the marginalized distribution is
\begin{equation}
    p_{B}(b)=\int_{\mathcal{A}} p_{A,B}(a,b) \mathrm{d} a. 
\end{equation} 
If the random variable is discrete instead of continuous, the integral is replaced with a summation.
The second basic operation is to form the \emph{conditional} distribution of one of the random variables, which is obtained by factorizing the joint distribution into a product of a conditional and a marginal distribution.
To form the conditional distribution of $A$ given $B$, the factorization is
\begin{equation}
    p_{A,B}(a,b) = p_{A \mid B}(a \mid b)p_{B}(b),
\end{equation}
where $p_{A \mid B}(a \mid b)$ is the sought-after conditional distribution.
The two basic operations above can be combined to form
\begin{align} \label{eq:Bayes}
    p_{B \mid A}(b \mid a)
        &= \frac{p_{A,B}(a,b)}{p_{A}(a)}
        = \frac{p_{A,B}(a,b)}{\int_{\mathcal{B}} p_{A,B}(a,b)\rmd b} \nonumber \\
        &= \frac{p_{A \mid B}(a \mid b)p_{B}(b)}{\int_{\mathcal{B}} p_{A \mid B}(a \mid b)p_{B}(b) \rmd b},
\end{align}
referred to as \textit{Bayes' theorem}.
\end{tcolorbox}

\begin{tcolorbox}[title=Example: Cascaded water tanks, breakable, enhanced, before upper={\parindent15pt\noindent}]
	The cascaded water tank system, depicted in \cref{fig:watertank}, consists of two water tanks placed vertically so that water can flow between them through holes in the bottom of the tanks.
A pump transports water from the reservoir below the tanks to the upper tank.
From the upper tank, the water can flow into the lower tank through a hole in the bottom of the upper tank.
Analogously, water can flow back into the reservoir through a hole in the bottom of the lower tank.
The inflow of water to the upper tank can be controlled by adjusting the voltage supplied to the pump, and the water level in the lower tank is measured.
The intended task is to control or predict the water level in the lower tank based on the voltage level supplied to the water pump.
The water flow in the system can, under normal operating conditions, be considered a weakly nonlinear process.
However, overflow can occur in both the upper and lower tank if the input signal is large enough, which corresponds to a hard nonlinearity.  

\subsubsection*{A state-space model of the cascaded water tanks based on physics}
The physical processes governing the behavior of fluid-flow systems like the water tanks are known and can be used to construct a discrete-time state-space model of the system.
\begin{center}
	\includegraphics[width=0.45\textwidth]{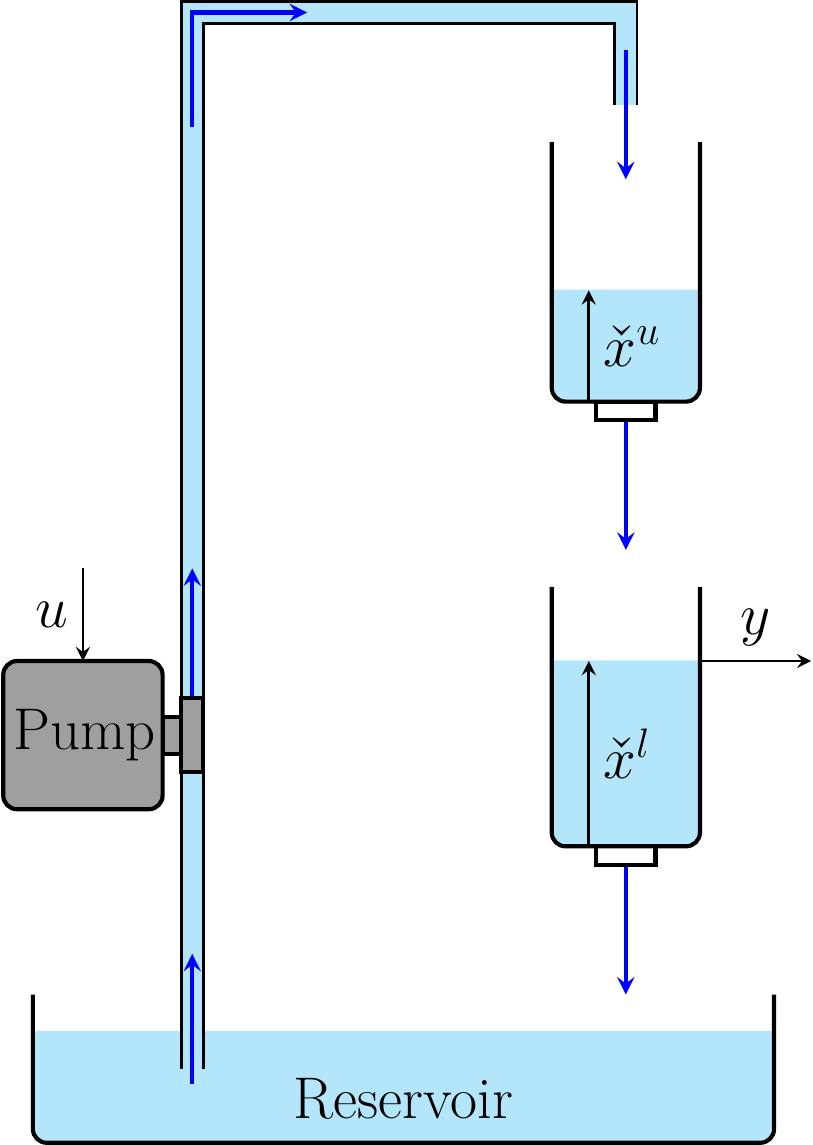}
\end{center}
\noindent \begin{minipage}{\textwidth}
	\captionof{figure}{
		The cascaded water tank system.
		The system consists of two water tanks placed vertically, so that water can flow between them.
		A pump transports water from the reservoir to the upper tank.
		From there, water flows into the lower tank through a hole in the bottom of the upper tank and back into the reservoir through a hole in the bottom of the lower tank.
		The inflow of water to the upper tank is controlled by adjusting the voltage $\inp$ supplied to the pump.
		The water level in the upper tank is ${\check{\state}}^u$, and the water level in the lower tank is ${\check{\state}}^l$.
		The observation $\obs$ is the measured water level in the lower tank.
		The water flow in the system is, under normal operating conditions, a weakly nonlinear process.
		However, both tanks can overflow if the input signal is large enough, which corresponds to a hard nonlinearity.
	}\label{fig:watertank}
\end{minipage}
One such model on functional form, taken from \cite{LindholmL:2019}, is
\begin{subequations}
	\label{eq:watertankSSM}
	\begin{align} 
	    \state_{\tv}^{u}
	        &= \check{\state}^{u}_{\tv-1} - k_1 \sqrt{\check{\state}^{u}_{\tv-1}} + k_5 \inp_{\tv} - k_2 \check{\state}^{u}_{\tv-1} + \pnoise_{\tv}^u, \label{eq:watertankSSM1} \\ 
	    \state_{\tv}^{l}
	            &= \check{\state}_{\tv-1}^{l} + k_1 \sqrt{\check{\state}_{\tv-1}^{u}} - k_3 \sqrt{\check{\state}_{\tv-1}^{l}} + k_2 \check{\state}_{\tv-1}^{u} - k_4 \check{\state}_{\tv-1}^{l} \nonumber \\
	            &\hspace{3.7mm}+ k_6 \max(0, \state_{\tv-1}^{u}-10) + \pnoise_{\tv}^{l} \label{eq:watertankSSM2}\\
	    \obs_{\tv} 
	        &= \check{\state}^{l}_{\tv} + \onoise_{\tv}, \label{eq:watertankSSM3}
	\end{align}
\end{subequations}
where the input $\inp_{\tv}$ is the voltage provided to the pump, and $\check{\state}_{\tv}^u$ and $\check{\state}_{\tv}^l$ are the water levels in the upper and lower tank, respectively (capped at 10 --- the height of the tanks).
The states of the model, $\state_{\tv}^u$ and $\state_{\tv}^l$, are the water levels in the upper and lower tank, respectively, plus the inflow to each tank.
The parameters $k_1, k_2, k_3, k_4, k_5, k_6$ encode physical properties of the system that are not known, for instance, the diameter of the tanks and holes, as well as different flow constants. 
The weakly nonlinear behavior of the tank system can be described using Bernoulli's principle and conservation of mass \cite{Schoukens2017}.
The terms in \cref{eq:watertankSSM1,eq:watertankSSM2} related to $k_1$, $k_3$, and $k_5$ capture this effect.
Bernoulli's principle is valid when losses in the system due to friction and the geometry of the tanks are low.
These losses are not negligible for the water tanks, so the linear terms $k_2 \check{\state}_{\tv}^{u}$ and $k_4 \check{\state}_{\tv}^{l}$ are added to the model to account for this effect \cite{Rogers2017}.
The hard nonlinearity introduced by overflow in the upper tank is accounted for through the term $k_6 \max(0, \state_{\tv-1}^{u} - 10)$, which represents an extra inflow to the lower tank when the water level in the upper tank is above $10$.
The noise terms $\pnoise_{\tv}^{u}, \pnoise_{\tv}^{l}, \onoise_{\tv}$ are assumed to be independent zero-mean, white Gaussian noise with unknown variances $\sigma_{\pnoise}^2$ and $\sigma_{\onoise}^2$.
The input voltage to the pump and the water level in the lower tank are measured.
The initial water level in the upper tank is unknown and must be estimated along with the parameters of the model, which are collectively referred to as $\vparam= \{ k_1, k_2, k_3, k_4, k_5, k_6, \sigma_{\pnoise}^2, \sigma_{\onoise}^2 \}$.

\subsubsection*{A probabilistic state-space model of the cascaded water tanks} 
All randomness in \cref{eq:watertankSSM} is due to the noise terms $\pnoise_{\tv}^u$, $\pnoise_{\tv}^l$, and $\onoise_{\tv}$, which are independent, additive, Gaussian random variables with mean value zero.
The probabilistic form of \cref{eq:watertankSSM} is therefore
\begin{subequations}
	\begin{align}
	    p(\vstate_{\tv} \mid \vstate_{\tv-1},\vparam)
	        &= \N{\bbm \mu_{\tv}^u(\state_{\tv-1}^u,\inp_{\tv}) \\ \mu_{\tv}^l(\state_{\tv-1}^u, \state_{\tv-1}^l) \ebm, \sigma_{\pnoise}^2 \I }, \\
	    p(\obs_{\tv} \mid \vstate_{\tv},\vparam)
	        &= \N{\check{\state}^{l}_{\tv}, \sigma_{\onoise}^2},
	\end{align}
\end{subequations}
where $\I$ is the identity matrix and the mean values are
\begin{subequations}
	\begin{align}
		\mu_{\tv}^u(\state_{\tv-1}^u,\inp_{\tv})
	        &= \check{\state}^{u}_{\tv-1} - k_1 \sqrt{\check{\state}^{u}_{\tv-1}} + k_5 \inp_{\tv} - k_2 \check{\state}^{u}_{\tv-1}, \\
	    \mu_{\tv}^l(\state_{\tv-1}^u, \state_{\tv-1}^l)
	        &= \check{\state}_{\tv-1}^{l} + k_1 \sqrt{\check{\state}_{\tv-1}^{u}} - k_3 \sqrt{\check{\state}_{\tv-1}^{l}} + k_2 \check{\state}_{\tv-1}^{u} \nonumber \\
	        &\hspace{3.8mm}- k_4 \check{\state}_{\tv-1}^{l} + k_{6} \max(0, \state_{\tv-1}^{u} \!-\! 10).
	\end{align}
\end{subequations}
The covariance matrix is diagonal, since the noise sources are assumed to be independent of each other.

\subsubsection*{The dataset}
For the cascaded water tank system in \cref{fig:watertank}, there exists a benchmark dataset developed at Uppsala University \cite{Schoukens2017} that consists of two data series: one for training and one for testing.
Both series have multisine input signals of length 1\thinspace024 with a frequency range of $0$ to $0.0144$~Hz.
The water level in the lower tank was measured using a capacitive water-level sensor.
Zeroth-order hold was used with a sampling period of $4$ s.
\end{tcolorbox}

\begin{tcolorbox}[title=Example: Dengue fever, breakable, enhanced, before upper={\parindent15pt\noindent}]
	Dengue fever is a tropical virus disease spread by mosquitoes that causes around 10,000 deaths every year worldwide \cite{GlobalBurden2016}. 
A human can become infected if bitten by an infected mosquito and, analogously, a mosquito can become infected if it bites an infected human.
There is no direct transmission of the virus between individuals of the same species.
After the initial bite, the virus goes into incubation (reproduction inside the host body without onset of any symptoms) for 8-12 days if the host is a mosquito or 4-10 days if the host is a human \cite{WHO}.
When the incubation period is over, the infected individual becomes infectious and can infect other individuals.
The infectious stage lasts 4-5 days on average, but it can be as long as 12 days for humans.
Mosquitoes are infectious for the rest of their life \cite{WHO}.
Humans either die from the virus, or recover and become immune after 2-7 days of symptoms \cite{WHO}.  

\subsubsection*{A state-space model for the spread of dengue fever based on prior knowledge} 
The dynamics of epidemiological diseases, like dengue fever or the coronavirus, can be described using \textit{compartmental models}.
In a compartmental model, the population is split into several disjoint compartments based on some disease-related characteristics (for example if an individual is infectious or have recovered from the disease).
A simple example of a compartmental model is the susceptible–infectious–recovered (SIR) model visualized in \cref{fig:SIR}.
It has three compartments: susceptible, which contains the individuals who are currently healthy but can become infected; infectious, which contains the currently infectious individuals; and recovered, which contains the individuals who have been sick but who are now either immune or dead.
In each instance, some proportion of the susceptible moves to the infectious compartment, and some proportion of the infectious move to the recovered compartment.
\begin{center}
	\includegraphics[width=.2\textwidth]{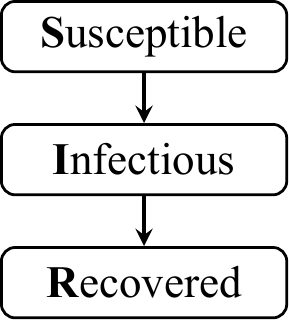}
\end{center}
\noindent \begin{minipage}{\textwidth}
	\captionof{figure}{
		The susceptible–infectious–recovered model is a compartmental model where the population is divided into three disjoint compartments: susceptible (healthy but can become infected), infectious (infectious and can infect others), and recovered (previously sick and is now either immune or dead).
		At each point in time, a proportion of the individuals in each compartment can transition from susceptible to infectious or from infectious to recovered.
		The transition is indicated by the arrows between the compartments.
	}\label{fig:SIR}
\end{minipage}	

To construct a model based on medical prior knowledge about the spread of dengue fever, two modifications to the SIR model are necessary.
Firstly, dengue fever has an incubation time, implying that individuals do not move directly from susceptible to infectious.
To take this into account, the compartment ``exposed'' is introduced, which contains individuals who have been infected but are not yet infectious.
Such a compartmental model is referred to as a susceptible–exposed-infectious–recovered (SEIR) model. Secondly, both humans and mosquitoes go through the stages of being susceptible $S(\ctv)$, exposed $E(\ctv)$, infectious $I(\ctv)$, and recovered $R(\ctv)$. However, they transition between compartments with different rates depending on, for instance, different incubation times.
Additionally, mosquitoes never enter the recovered compartment.
To incorporate the different dynamics for humans and mosquitoes, two coupled SEIR models are used\textemdash one for humans and one for mosquitoes.
The final, coupled SEIR model is depicted in \cref{fig:SEIRdengue}.
\begin{center}
	\includegraphics[width=.75\textwidth]{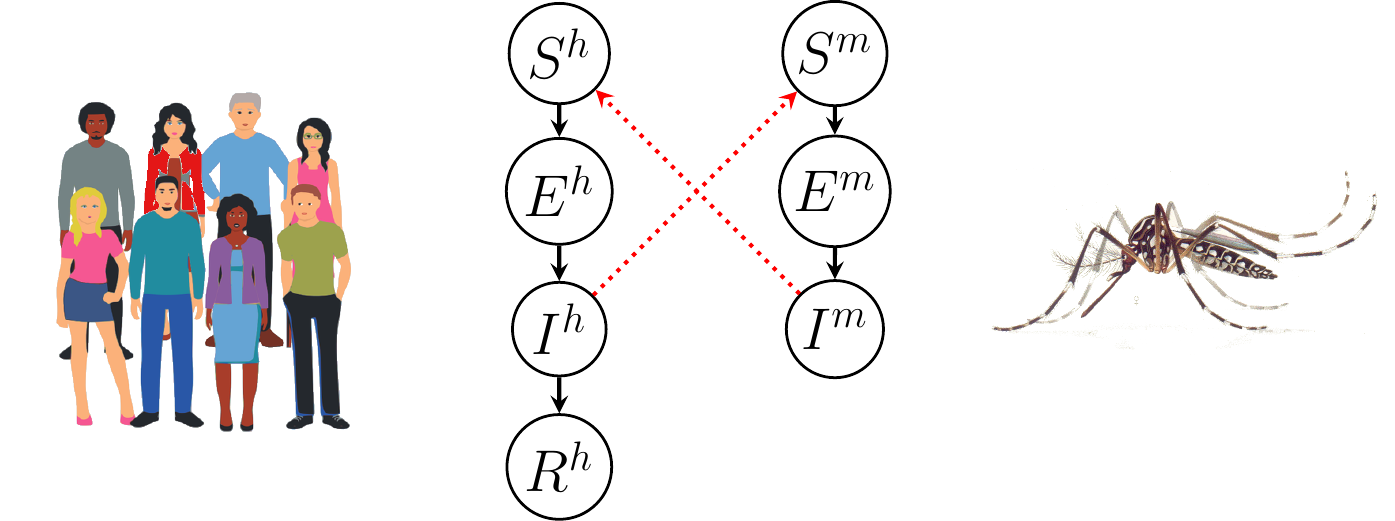}
\end{center}
\noindent \begin{minipage}{\textwidth}
	\captionof{figure}{
		A coupled susceptible–exposed-infectious–recovered (SEIR) model for an outbreak of dengue fever.
		To model the spread of the disease, two coupled SEIR models are used to capture that infectious mosquitoes can infect humans and infectious humans can infect mosquitoes. The coupling is illustrated with the red, dotted arrows between the two SEIR models.
		The SEIR model is a compartmental model with four compartments: susceptible, exposed, infectious, and recovered.
		The black, solid arrows indicate transitions between compartments.
		The difference, compared to the SIR model in \cref{fig:SIR}, is that the exposed compartment has been added to account for the incubation period of the dengue virus.
		The model for the mosquitoes lack the recovered compartment, since mosquitoes die before they recover from the disease.
	}\label{fig:SEIRdengue}
\end{minipage}

A continuous-time state-space model corresponding to \cref{fig:SEIRdengue} can be based on the system of coupled, ordinary differential equations (ODEs)
\begin{equation}
\label{eq:SebODE}
\begin{aligned} 
	\frac{\rmd S^{h}}{\rmd \ctv}(\ctv) &=-\lambda^{h}S^{h}(\ctv),    					  &  \frac{\rmd S^{m}}{\rmd \ctv}(\ctv) &= \nu^{m}-(\lambda^{m} + \mu^{m}) S^{m}(\ctv),  \\
	\frac{\rmd E^{h}}{\rmd \ctv}(\ctv) &= \lambda^{h}S^{h}(\ctv) - \delta^{h}E^{h}(\ctv), &  \frac{dE^{m}}{\rmd \ctv}(\ctv) &= \lambda^{m}S^{m}(\ctv) - (\delta^{m} + \mu^{m}) E^{m}(\ctv),  \\
	\frac{\rmd I^{h}}{\rmd \ctv}(\ctv) &= \delta^{h}E^{h}(\ctv) - \gamma^{h} I^{h}(\ctv),&  \frac{\rmd I^{m}}{\rmd \ctv}(\ctv) &= \delta^{m}E^{m}(\ctv) - \mu^{m} I^{m}(\ctv),  \\
	\frac{\rmd R^{h}}{\rmd \ctv}(\ctv) &= \gamma^{h}I^{h}(\ctv),%
\end{aligned}
\end{equation}
as described in \cite{Funk16}.
The state is $\vstate = [S^{h}, E^{h}, I^{h}, R^{h}, S^{m}, E^{m}, I^{m}, R^{m}]^+$, and the parameters are the force of infection $\lambda$, the incubation rate $\delta$, the recovery rate $\gamma$, the birth rate $\nu$, and the death rate $\mu$.
The superscripts $h$ and $m$ denote human and mosquito, respectively.
The duration of the outbreak is assumed to be short enough to make human births and deaths negligible.
The coupling between the two SEIR models stems from the force of infection parameters, which are modeled as $\lambda^h=c_hI^m$ and $\lambda^m=c_mI^h$, where the exact form of the constants $c_h$ and $c_m$ are available in \cite{Funk16}.
The observations are Poisson-distributed with a rate that depends on the number of newly infectious humans and the reporting rate $\rho$ to health centers.

\subsubsection*{A probabilistic state-space model of the spread of dengue fever}
The continuous-time state-space model in \cref{eq:SebODE} describes the dynamics using ODE.
For compartmental models, this corresponds to a mean-field approximation of an underlying stochastic process.
Such an approximation is valid for large populations. However, it can be inaccurate for small populations, especially if the number of infectious drops below one \cite{Kurtz1970}.
A probabilistic, discrete-time, and discrete-state version of \cref{eq:SebODE} was derived in \cite{DelaySampling2018}.
The state transitions between compartments are now stochastic and are performed in two steps\textemdash a transition that keeps the population fixed, followed by a transition that accounts for births and deaths.
Only the former transition is discussed here.
The transitions between compartments (excluding births and deaths) are
\begin{equation}
\label{eq:LawProb}
\begin{aligned} 
	S^{h}_{\tv} &= S^{h}_{\tv-1} - e^{h}_{\tv},   			 & S^{m}_{\tv} &= S^{m}_{\tv-1}-e^{m}_{\tv},  \\
	E^{h}_{\tv} &= E^{h}_{\tv-1}+e^{h}_{\tv}-i^{h}_{\tv},	 	 & E^{m}_{\tv} &= E^{m}_{\tv-1} + e^{m}_{\tv} - i^{m}_{\tv}, \\
	I^{h}_{\tv} &= I^{h}_{\tv-1} + i^{h}_{\tv} - r^{h}_{\tv}, 	 & I^{m}_{\tv} &= I^{m}_{\tv-1} + i^{m}_{\tv} - r^{m}_{\tv},  \\
	R^{h}_{\tv} &= R^{h}_{\tv-1} + r^{h}_{\tv}, 
\end{aligned}
\end{equation}
where lowercase letters indicate the stochastic number of newly exposed $e_{\tv}$, infectious $i_{\tv}$, and recovered $r_{\tv}$.
These random variables are assumed to be binomially-distributed.
The binomial distribution, $\text{Bin}(n,p)$, describes the number of successes in $n$ independent trials with only two possible outcomes, success or failure, where success occurs with probability $p$.
The number of newly infectious humans in an SEIR model can be modeled as the number of (unfortunate) successful transitions from the exposed compartment to the infectious compartment.
The number of exposed individuals that can transition at time $\tv$ is $E_{\tv-1}^h$, and a transition of one individual occurs with an infectious probability $\delta^h$, independent of all other exposed individuals.
Thus, the number of newly infectious is a binomial variable $i_{\tv}^h \sim \text{Bin}(E_{\tv-1}^h, \delta^h)$.
The transitions between all compartments are
\begin{equation}
\label{eq:DengueStateUpdate}
\begin{aligned} 
	e^{h}_{\tv} &\sim \text{Bin}(e^{h}_{\tv} \mid \tau^{h}_{\tv}, \lambda^{h}), & e^{m}_{\tv} &\sim \text{Bin}(e^{m}_{\tv} \mid \tau^{m}_{\tv}, \lambda^{m}),  \\
	i^{h}_{\tv} &\sim \text{Bin}(i^{h}_{\tv} \mid E^{h}_{\tv-1}, \delta^{h}),   & i^{m}_{\tv} &\sim \text{Bin}(i^{m}_{\tv} \mid E^{m}_{\tv-1}, \delta^{m}),  \\
	r^{h}_{\tv} &\sim \text{Bin}(r^{h}_{\tv} \mid I^{h}_{\tv-1}, \gamma^{h}),   & r^{m}_{\tv} &\sim \text{Bin}(r^{m}_{\tv} \mid I^{m}_{\tv-1}, \gamma^{m}), 
\end{aligned}
\end{equation}
where the parameters are the transmission probability $\lambda$, the infection probability $\delta$, and the recovery probability $\gamma$. 
Mosquitoes never recover from the disease. Thus, $\gamma^m$ is always zero, ensuring that $r^{m}_{\tv}$ is zero.
The number of susceptible humans that were bitten by an infectious mosquito, $\tau_{\tv}^h$, and the number of susceptible mosquitoes that have bitten an infectious human, $\tau_{\tv}^m$, are
\begin{equation}
\label{eq:DengueTao}
\begin{aligned}
	\tau^{h}_{\tv} &\sim \text{Bin}\left(S_{\tv-1}^{h}, 1-\exp\left(-I_{\tv-1}^{m}/n_{\tv-1}^{h} \right)\right),  \\
\tau^{m}_{\tv} &\sim \text{Bin}\left(S_{\tv-1}^{m}, 1-\exp\left(-I_{\tv-1}^{h}/n_{\tv-1}^{h} \right)\right). 
\end{aligned}
\end{equation}
The derivation of these quantities can be found in \cite{DelaySampling2018}.
The observations are the number of newly infectious humans that sought medical care at a health center.
Only a proportion of all that have been infected are expected to report to health centers. Thus, the observations are also modeled as binomially-distributed with reporting probability $\rho$, that is, 
\begin{equation} \label{eq:dengueObs}
	\obs_{\tv} \sim \text{Bin}(i_{\tv}^h,\rho).
\end{equation}
The complete parameter vector is $\vparam = \{ \lambda^h, \delta^h, \gamma^h, \lambda^m, \delta^m, \gamma^m, \rho \}$ for this model.

\subsubsection*{The dataset}
The dataset contains $197$ observations of the number of newly infectious humans that visited a health center during an outbreak of dengue fever on the island Yap in Micronesia in $2011$.
The cases were reported daily during the main outbreak and weekly before and after.
In total, there were $978$ reported cases in a population of $7\thinspace370$.
The dataset was originally presented in \cite{Funk16}.
\end{tcolorbox}


\section{Identification}
Identification of the state-space model \cref{eq:state-space model} is the process of learning all the unknown parameters so that the model best describes some measured input-output data.
Let $\data_{T} =  \{\vobs_{\tv}, \vinp_{\tv}\}_{\tv=1}^{T}$ denote the collection of measured outputs and possible inputs up to time $T$. 
More precisely, the focus in this article is on the \emph{batch} system identification problem, which amounts to finding a description for the unknown parameters $\vparam$ of the state-space model \cref{eq:state-space model} based on the available data $\data_{T}$.
For notational simplicity\textemdash without loss of generality\textemdash the known input $\vinp_{\tv}$ is from now on dropped from the notation. 

To proceed with identification, it is important to acknowledge any assumptions, or prior knowledge, on the unknown parameter values themselves.
The two most commonly used assumptions on the unknown parameters $\vparam$ are:
\begin{enumerate}
    \item 
        \textit{Frequentistic (ML):}
        The parameters are assumed to be deterministic variables.
        The aim is to find a point estimate of the parameters.
    \item
        \textit{Bayesian:}
        The parameters are assumed to be random variables, implying that the model must be augmented with a prior distribution $\vparam \sim p(\vparam)$ for the parameters.
        The aim is to find the posterior distribution of the parameters.
\end{enumerate}
Both formulations are treated in this article, without making any individual ranking between them.
One of these may be better suited for a given problem, or perhaps a combination makes most sense\textemdash it depends on the task at hand.
Which assumption to use is an important decision, since it influences which identification algorithms that can be applied. 

ML is intuitive in the sense that it amounts to finding the point estimate of the unknown parameters that makes the observed data as likely as possible.
This is done by selecting the parameter value that maximizes the marginal likelihood $p(\vobs_{\range{1}{T}} \mid \vparam)$ of the observed data $\vobs_{\range{1}{T}} = (\vobs_1, \vobs_2, \dots, \vobs_T)$,
\begin{align}
    \what{\vparam}_{\text{ML}}
        &= \argmax_{\vparam} p(\vobs_{\range{1}{T}} \mid \vparam).
    \label{qe:maximum-likelihood}
\end{align}
ML is one of the most common ways to formulate system identification problems.
It has been extensively covered in the literature, see e.g. \cite{Ljung:2000,SoderstromS:1989}.

In the Bayesian formulation, the aim is instead to find the posterior distribution of the parameters $p(\vparam\mid \vobs_{\range{1}{T}})$.
From an application of Bayes' theorem  \eqref{eq:Bayes}, the parameter posterior is
\begin{align}
    \label{eq:Bayesian}
    p(\vparam \mid \vobs_{\range{1}{T}}) = \frac{p(\vobs_{\range{1}{T}} \mid \vparam) p(\vparam)}{p(\vobs_{\range{1}{T}})}.
\end{align}
Bayesian system identification is not as well-developed as its frequentistic counterparts. However, there has recently been more developments along this line of research. See \cite{Peterka:1981} for early work and \cite{AndrieuDH:2010,SchonSML:2018} for newer initiatives.

It is interesting to note that the likelihood $p(\vobs_{\range{1}{T}} \mid \vparam)$ is required for both the ML \eqref{qe:maximum-likelihood} and the Bayesian \cref{eq:Bayesian} formulation. 
For a state-space model, the  likelihood can be computed via the integral
\begin{align}
    \label{eq:MargLikelihood}
    p(\vobs_{\range{1}{T}} \mid \vparam) = \int p(\vobs_{\range{1}{T}}, \vstate_{\range{1}{T}} \mid \vparam) \;\rmd \vstate_{\range{1}{T}},
\end{align}
that is, by marginalizing out the hidden sequence of state variables $\vstate_{\range{1}{T}}$.
This illustrates a central aspect of working with state-space models.
In addition to the model parameters and the observed data, the model also contains hidden (unobserved) state variables that must be handled in some way, typically by marginalization.
Indeed, the integrand in the expression above is the full probabilistic model for which there exists a tractable expression,
\begin{align}
\label{eq:FullLikelihood}
    p(\vobs_{\range{1}{T}}, \vstate_{\range{1}{T}} \mid \vparam) = p(\vstate_{1} \mid \vparam) \prod_{\tv=1}^{T} p(\vobs_{\tv} \mid \vstate_{\tv}, \vparam) \prod_{\tv=2}^{T} p(\vstate_{\tv} \mid \vstate_{\tv-1},\vparam).
\end{align}
However, computing the integral is challenging in general.

One interpretation of the likelihood calculation \cref{eq:MargLikelihood} is that it amounts to averaging the full probabilistic model $p(\vobs_{\range{1}{T}}, \vstate_{\range{1}{T}} \mid \vparam)$ over all possible state trajectories $\vstate_{\range{1}{T}}$.
For nonlinear system identification, the need for approximations \textemdash like the ones offered by SMC methods \textemdash is tightly linked to the intractability of \cref{eq:MargLikelihood} and the unknown state trajectory $\vstate_{\range{1}{T}}$.
The two main strategies for addressing the unknown state trajectory are: 
\begin{enumerate}
    \item
        \textit{Marginalization:}
        In this strategy, the state variables are marginalized (integrated out) according to \cref{eq:MargLikelihood}, implying that the parameters are targeted directly.
        The identification problem is solved by first computing the integral appearing in \cref{eq:MargLikelihood}, continuing by viewing $\vparam$ as the only unknown quantity of interest. 
        In the frequentistic problem formulation, the prediction error method and direct maximization of the likelihood belong to this strategy \cite{Ljung:2000}.
        In the Bayesian formulation, the Metropolis--Hasitngs (MH) algorithm \cite{Metropolis:1953}-\cite{Hastings:1970} can be used to approximate the posterior distribution of the parameters conditioned on the data.
    \item
        \textit{Data augmentation:}
        In this strategy, the states are treated as auxiliary variables that are estimated together with the parameters.
        Intuitively, this strategy can be thought of as a systematic way of separating \emph{one} hard problem into \emph{two} new and closely linked subproblems, each of which should be easier to solve than the original problem.
        The expectation maximization (EM) algorithm \cite{DempsterLR:1977} solves the ML formulation in this way, and the Gibbs sampler \cite{GemanG:1984} solves the Bayesian problem using this strategy.
\end{enumerate}

The important difference between these strategies is that marginalization targets the parameters only, whereas data augmentation targets both parameters and states.
Thus, marginalization, operates on a smaller dimensional space, but must somehow handle the intractable likelihood.
On the other hand, data augmentation operates on a much larger space, but the (complete) likelihood \cref{eq:FullLikelihood} is available on closed form.

\begin{figure}[htb]
    \centering 
    \includegraphics[width=.65\textwidth]{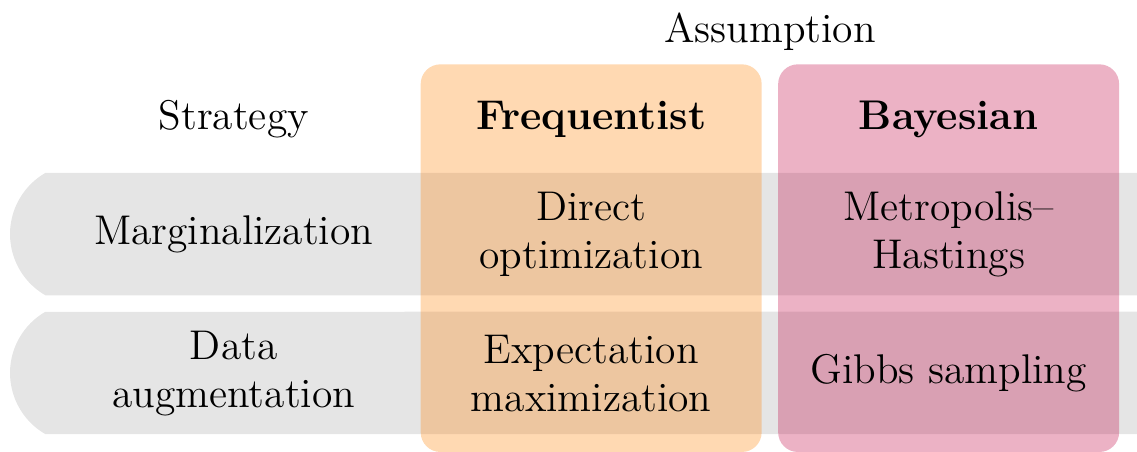}
    \caption{
        The structuring of approaches used for system identification in this article.
        The rows divide the methods into two strategies (marginalization and data augmentation), whereas the columns describe two different assumptions on how to treat the unknown parameters (frequentistic and Bayesian). 
    }
    \label{fig:SIoverview}
\end{figure}

In this article, the identification methods have been associated with one of the two strategies based on where they fit most naturally.
This division is visualized in \cref{fig:SIoverview}.
For instance, Gibbs sampling alternates between sampling the parameters and sampling the states; hence, it fits well with the data augmentation strategy.
MH, on the other hand, can sample the parameters directly and fits more naturally with the marginalization strategy. 
It is important to note that many identification algorithms can be used for both strategies, even if they are better-suited for one of them.
For example, MH could also be applied for the data augmentation strategy by sampling parameters and states jointly.
However, it is well-known that MH can be inefficient for high-dimensional problems. Gibbs sampling, for example, typically performs better in this case. 

Regardless of whether a marginalization or data augmentation approach is adopted, the marginal likelihood $p(\vobs_{\range{1}{T}} \mid \vparam)$ plays a central role.
While it is generally intractable to evaluate exactly, the next section discusses a class of SMC methods that offer tractable approximations of expectation integrals like \cref{eq:MargLikelihood}.
SMC forms the basis for the ensuing algorithms that employ either the maximum-likelihood or Bayesian assumption on the parameters, and either the marginalization or data augmentation strategy to address the unknown state trajectory.


\section{Sequential Monte Carlo}
SMC methods \cite{Gordon:1993,Kitagawa:1993,StewartM:1992} are generic algorithms for performing approximate inference in statistical models.
This section first gives a brief description of the general SMC framework and then specializes to nonlinear state-space models. 

SMC is based on importance sampling and uses weighted samples to iteratively approximate a sequence of probability distributions, referred to as the \textit{target distributions}.
The target distribution at iteration $\tv$, denoted $\target_{\tv}(\vstate_{\range{1}{\tv}})$, is a joint probability density function (pdf) over the latent variables $\vstate_{\range{1}{\tv}} = (\vstate_1, \dots, \vstate_{\tv})$.
The pdf is
\begin{align}
    \target_{\tv}(\vstate_{\range{1}{\tv}}) = \frac{1}{\Z_{\tv}} \utarget_{\tv}(\vstate_{\range{1}{\tv}}), \qquad \tv=1,\dots, T,
    \label{eq:target}
\end{align}
where $\utarget_{\tv}$ is a positive integrable function that can be evaluated point-wise, and the normalization constant $\Z_{\tv} = \int \utarget_{\tv}(\vstate_{\range{1}{\tv}}) \;\rmd \vstate_{\range{1}{\tv}}$ ensures that $\target_{\tv}$ is a pdf.

SMC approximates each target $\target_{\tv}$ by a collection of $\Npar$ weighted particles $\{\vstate_{\range{1}{\tv}}^{\npar}, \w_{\tv}^{\npar}\}_{\npar=1}^{\Npar}$ that are generated according to \cref{alg:sequential-monte-carlo}.

\begin{algorithm}[htb]
    \begin{algorithmic}
    	\State Sample $\vstate_{1}^{\npar} \sim \q_{1}(\vstate_1)$ independently, compute unnormalized weights $\uw_{1}^{\npar} = \utarget_{1} (\vstate_{1}^{\npar})/\q_{1}(\vstate_{1}^{\npar})$ and normalize $\w_{1}^{\npar} = \uw_{1}^{\npar} / \sum_{j=1}^{\Npar} \uw_{1}^{j}$.
    	\For{$\tv=2,\dots,T$}
    		\State (a) \textit{Resample:} Simulate ancestor indices $a_{\tv}^i$ with probabilities $\{\w_{\tv-1}^{i}\}_{i=1}^N$.
    		\State (b) \textit{Propagate:} Simulate $\vstate_{\tv}^{i} \sim \q_{\tv}(\vstate_{\tv} \mid \vstate_{\tv-1}^{a_{\tv}^{i}})$ and set $\vstate_{\range{1}{\tv}}^{i} = \{ \vstate_{\range{1}{\tv-1}}^{a_{\tv}^{i}}, \vstate_{\tv}^{i} \}$.
    		\State (c) \textit{Weight:} Compute $\uw_{\tv}^{i} = \w_{\tv}(\vstate_{\range{1}{\tv}}^{i})$ and normalize $\w_{\tv}^{i} = \uw_{\tv}^{i} / \sum_{j=1}^{N} \uw_{\tv}^{j}$.
    		\EndFor
    \end{algorithmic}
    \caption{Sequential Monte Carlo (all steps are for $i=1,\ldots,N$)}
    \label{alg:sequential-monte-carlo}
\end{algorithm}
The algorithm initializes each particle by simulating independent samples from a user-specified proposal distribution $\q_{1}(\vstate_{1})$ and computing the normalized weights $\{\w_{1}^{i}\}_{i=1}^{N}$.
How to choose the proposal distribution will be discussed later.
For each iteration $\tv$, the particles are resampled by simulating ancestor indices $a_{\tv}^{i}$ for each particle, with probabilities given by the normalized weights $\{\w_{\tv-1}\}_ {i=1}^{N}$.
This produces an unweighted approximation of $\target_{\tv-1}$ by creating replicates of particles with high weights and discarding particles with low weights.
Given this unweighted approximation, the particles are propagated by simulating the proposal $\q_{\tv}(\vstate_{\tv} \mid \vstate_{\tv-1})$ for each particle and computing new, unnormalized weights according to the weighting function
\begin{align}
    \w_{\tv}(\vstate_{\range{1}{\tv}}) = \frac{\utarget_{\tv}(\vstate_{\range{1}{\tv}})}{\utarget_{\tv-1}(\vstate_{\range{1}{\tv-1}}) \q_{\tv}(\vstate_{\tv} \mid \vstate_{\tv-1})}.
    \label{eq:weight-function}
\end{align}
The proposal can be chosen arbitrarily, as long as its support includes the support of the target $\target_{\tv}(\vstate_{\tv} \mid \vstate_{\range{1}{\tv-1}})$.
One possible choice is the \textit{locally optimal proposal},  
\begin{align}
    \q_{\tv}(\vstate_{\tv} \mid \vstate_{\tv-1}) \propto \frac{\utarget_{\tv}(\vstate_{\range{1}{\tv}})}{\utarget_{\tv-1}(\vstate_{\range{1}{\tv-1}})},
\end{align}
which minimizes the conditional variance of the weights at iteration~$\tv$.
The weighted particles generated by \cref{alg:sequential-monte-carlo} can be used to approximate each intermediate target distribution $\target_{\tv}$ by the empirical distribution $\widehat{\target}_{\tv}(\vstate_{\range{1}{\tv}}) = \sum_{i=1}^{N} \w_{\tv}^{i} \delta_{\vstate_{\range{1}{\tv}}^{i}}(\vstate_{\range{1}{\tv}})$.
The algorithm also provides \emph{unbiased} estimates of the normalizing constants $\Z_\tv$, \begin{align} \label{eq:SMClikelihood}
    \widehat{\Z_{\tv}} = \prod_{s=1}^{\tv} \frac{1}{N} \sum_{i=1}^{N} \uw_{s}^{i},
\end{align}
see \cite{del2004feynman} and \cite{PittSGK:2012} for details. Textbook introductions to SMC are provided in \cite{ChopinP:2020,Sarkka:2013,CappeMR:2005}.

\subsection{Designing the target for state-space models}
To enable inference in state-space models, the target distributions must be connected to a relevant sequence of distributions.
Analog to \cref{eq:FullLikelihood}, the joint pdf of the latent states and observations up to time $\tv$ is
\begin{align}
    \label{eq:smc:joint-filter-target}
    p(\vobs_{\range{1}{\tv}}, \vstate_{\range{1}{\tv}})
        &= p(\vstate_{1}) \prod_{s=1}^{\tv} p(\vobs_{s} \mid \vstate_{s}) \prod_{s=2}^{\tv} p(\vstate_{s} \mid \vstate_{s-1}),
\end{align}
where the dependence on the parameters has been dropped to simplify the notation.
Letting the unnormalized target distributions be  $\utarget_{\tv}(\vstate_{\range{1}{\tv}}) = p( \vobs_{\range{1}{\tv}}, \vstate_{\range{1}{\tv}})$ for $\tv=1,\ldots,T$ is therefore a natural choice that links the sequential structure of the algorithm with the sequential nature of the model.  
Indeed, this link has driven the development of SMC methods and is the foundation for the original particle filter (which is a special case of SMC).
See ``Background: The bootstrap particle filter'' for further details and an alternative derivation of the particle filter.
The targets $\utarget_{\tv}(\vstate_{\range{1}{\tv}}) = p( \vobs_{\range{1}{\tv}}, \vstate_{\range{1}{\tv}})$ recursively estimates the joint filtering distribution $p(\vstate_{\range{1}{\tv}} \mid \vobs_{\range{1}{\tv}})$ and produces an unbiased estimate of the marginal likelihood $\Z_{\tv} = p(\vobs_{\range{1}{\tv}})$.

In offline system identification problems, it is often not important to estimate the filtering distributions.
In fact, in most cases, only the final target $\target_{T}(\vstate_{\range{1}{T}})$ is of interest.
For the state-space model, this corresponds to the final filtering distribution, which equals the smoothing distribution, namely the distribution of all states given all observations $p(\vstate_{\range{1}{T}} \mid \vobs_{\range{1}{T}})$.
Requiring only $\target_{T}(\vstate_{\range{1}{T}}) = p(\vstate_{\range{1}{T}} \mid \vobs_{\range{1}{T}})$ gives a \emph{larger freedom} in designing the sequence leading up to the final target.
This freedom can be used to improve the final estimate, both for estimating the marginal likelihood and approximating the smoothing distribution.

A problem with using the filtering distributions as target distributions is that the support for the state $\vstate_{\tv}$ at iteration $\tv$ is determined by the filtering approximation at iteration $\tv$.
The marginals of the filtering distribution $p(\vstate_{\tv} \mid \vobs_{\range{1}{\tv}})$ and the marginal smoothing distribution $p(\vstate_{\tv} \mid \vobs_{\range{1}{T}})$ can differ a lot depending on the model and the observations.
When the SMC algorithm propagates the particles from iteration $\tv-1$ to $\tv$, the support for the states $\vstate_{\range{1}{\tv-1}}$ does not change.
The new state $\vstate_{\tv}$ has a distribution determined by the model, but $\vstate_{\range{1}{\tv-1}}$ has finite support.
This leads to a situation where eventually all particles share a common ancestor and the marginal distribution of the states before that time is represented by a single sample.
This is referred to as \emph{degeneracy}.

The freedom in designing the intermediate target distributions leading up to the final target raises the question of an optimal sequence.
For estimating the normalizing constant $\Z_{T}$, the optimal target distributions are marginals of the final target
\begin{align}
    \target_{\tv}(\vstate_{\range{1}{\tv}})
        = \int \target_{T}(\vstate_{\range{1}{T}}) \;\rmd \vstate_{\range{\tv+1}{T}}
        = \target_{T}(\vstate_{\range{1}{\tv}}).
        \label{eq:marignal-targets}
\end{align}
The locally optimal proposal for the optimal target sequence is
\begin{align}
    \q^{\star}_{\tv}(\vstate_{\tv} \mid \vstate_{\tv-1})
        &\propto \frac{\utarget_{\tv}(\vstate_{\range{1}{\tv}})}{\utarget_{\tv-1}(\vstate_{\range{1}{\tv-1}})}
        = \frac{\utarget_{T}(\vstate_{\range{1}{\tv}})}{\utarget_{T}(\vstate_{\range{1}{\tv-1}})} = \target_{T}(\vstate_{\tv} \mid \vstate_{\range{1}{\tv-1}})
        = p(\vstate_{\tv} \mid \vstate_{{\tv-1}}, \vobs_{\range{\tv}{T}}),
        \label{eq:optimal-proposal}
\end{align}
where the last equality is true for the state-space model.
This proposal is typically not possible to simulate, since it requires exact samples from the posterior distribution.
However, using marginals of the final distribution as target distributions in combination with the locally optimal proposal leads to $\what{\Z_{\tv}} = \Z_{T}$ with probability 1 for all $\tv$, since
\begin{align}
    \w_{1}(\vstate_1)
        &= \frac{\utarget_{1}(\vstate_{1})}{\q_{1}(\vstate_1)}
         = \frac{\utarget_{1}(\vstate_{1})}{\utarget_{1}(\vstate_{1})/\int \utarget_{1}(\vstate_{1}) \;\rmd \vstate_{1}}
        = \iint \utarget_{T}(\vstate_{\range{1}{T}}) \;\rmd \vstate_{\range{2}{T}} \rmd \vstate_{1}
         = \Z_T,
\end{align}
and $\w_{\tv}(\vstate_{\range{1}{\tv}}) = {\utarget_{\tv}(\vstate_{\range{1}{\tv}})}/({\utarget_{\tv-1}(\vstate_{\range{1}{\tv-1}}) \q_{\tv}(\vstate_{\tv} \mid \vstate_{\tv-1})}) \equiv 1$ for $\tv = 2,\ldots,T$.

For the optimal target, both the weight expression \cref{eq:weight-function} and sampling or evaluating the proposal are intractable.
It does, however, give useful guidance for the design of both the intermediate targets and the proposal. 
For a state-space model, the weight expression is
\begin{align}
    \w_{\tv}(\vstate_{\range{1}{\tv}})
        &= \frac{p(\vstate_{\range{1}{\tv}}, \vobs_{\range{1}{T}})}{p(\vstate_{\range{1}{\tv-1}}, \vobs_{\range{1}{T}}) \q_{\tv}(\vstate_{\tv} \mid \vstate_{\tv-1})}
        = \frac{p(\vstate_{\tv} \mid \vstate_{\tv-1}) p(\vobs_{\tv} \mid \vstate_{\tv}) p(\vobs_{\range{\tv+1}{T}} \mid \vstate_{\tv})}{\q_{\tv}(\vstate_{\tv} \mid \vstate_{\tv-1}) p(\vobs_{\range{\tv}{T}} \mid \vstate_{\tv-1})}.
        \label{eq:twisted_weights}
\end{align}
Let $\twist_{\tv}(\vstate_{\tv}) = p(\vobs_{\range{\tv+1}{T}} \mid \vstate_{\tv})$ and $\twist_{0} \equiv \twist_{T} \equiv 1$.
The functions $\twist_{\tv}$ are called the \emph{twisting potentials} \cite{whiteley2014twisted,heng2020controlled,guarniero2017iterated}.
Evaluating the weight expression \cref{eq:twisted_weights} is intractable for most models, since it would require evaluation of the conditional marginal likelihood.
However, since the twisting potential for the final target $\twist_{T}$ is constant, the choice of twisting potentials does not alter the final target (only the sequence leading up to it).
From \cref{eq:twisted_weights}, the optimal twisting potential for state-space models is $\twist^{\star}_{\tv}(\vstate_{\tv}) = p(\vobs_{\range{\tv+1}{T}} \mid \vstate_{\tv})$.
One possibility is to approximate the conditional likelihood using deterministic inference methods such as the extended or the unscented Kalman filter \cite{sarkka2013bayesian}.
It is also possible to use an extended Rauch--Tung--Striebel or unscented smoother to construct the proposal from a deterministic approximation.
An example where the likelihood $p(\vobs_{\range{1}{T}} \mid \vparam)$ is estimated using both a bootstrap particle filter and a twisted particle filter is available in ``Example: Twisted particle filter for the cascaded water tanks''.

In this way, SMC can be seen as running on top of a biased inference method known to work well while also providing statistical guarantees such as unbiased estimates of the marginal likelihood.

\begin{tcolorbox}[title=Background: The bootstrap particle filter ,breakable, enhanced, before upper={\parindent15pt\noindent}]
	Sequential Monte Carlo (SMC) methods were initially known as particle filters \cite{Gordon:1993,Kitagawa:1993,StewartM:1992} and presented as an approximate solution to the optimal filtering problem for nonlinear state-space models.
In this section, the SMC method is introduced via optimal filtering in state-space models.

The filtering problem amounts to computing the marginal filtering distribution $p(\vstate_{\tv} \mid \vobs_{\range{1}{\tv}})$ sequentially in time for $\tv = 1, \ldots, T$.
By applying the laws of conditional distributions, the filtering distribution can be written
\begin{align}
    p(\vstate_{\tv} \mid \vobs_{\range{1}{\tv}})
        &= \frac{p(\vobs_{\tv} \mid \vstate_{\tv}) p(\vstate_{\tv} \mid \vobs_{\range{1}{\tv-1}})}{p(\vobs_{\tv} \mid \vobs_{\range{1}{\tv-1}})},
        \label{eq:update-step}
\end{align}
where the denominator is $p(\vobs_{\tv} \mid \vobs_{\range{1}{\tv-1}}) = \int p(\vobs_{\tv} \mid \vstate_{\tv}) p(\vstate_{\tv} \mid \vobs_{\range{1}{\tv-1}}) \;\rmd \vstate_{\tv}$.
The expression \eqref{eq:update-step} is often referred to as the update step, since it updates the predictive distribution $p(\vstate_{\tv} \mid \vobs_{\range{1}{\tv-1}})$ with the most recent observation $\vobs_{\tv}$.
Using marginalization, the predictive distribution is
\begin{align}
    p(\vstate_{\tv} \mid \vobs_{\range{1}{\tv-1}})
        &= \int p(\vstate_{\tv} \mid \vstate_{\tv-1}) p(\vstate_{\tv-1} \mid \vobs_{\range{1}{\tv-1}}) \;\rmd \vstate_{\tv-1},
        \label{eq:prediction-step}
\end{align}
where the second term in the product is the filtering distribution at time $\tv-1$.
This is referred to as the prediction step, since it predicts the next state given the current filtering distribution.

Both \cref{eq:update-step,eq:prediction-step} are typically intractable.
Particle filters approximate these distributions using weighted samples.
Assume that the filtering density at time $\tv-1$ has been approximated by a set of $N$ weighted samples $\{\w_{\tv-1}^i, \vstate_{\tv-1}^{i}\}_{i=1}^{N}$ as $\what{p}(\vstate_{\tv-1} \mid \vobs_{\range{1}{\tv-1}}) = \sum_{i=1}^{N} \w_{\tv-1}^{i} \delta_{\state_{\tv-1}^i}(\vstate_{\tv-1})$, where $\delta_x$ is the Dirac delta.
The prediction step \cref{eq:prediction-step} then reduces to
\begin{align}
    p(\vstate_{\tv} \mid \vobs_{\range{1}{\tv-1}})
        &\approx \int p(\vstate_{\tv} \mid \vstate_{\tv-1}) \what{p}(\vstate_{\tv-1} \mid \vobs_{\range{1}{\tv-1}}) \;\rmd \vstate_{\tv-1} = \sum_{i=1}^{N} \w_{\tv-1}^{i} p(\vstate_{\tv} \mid \vstate_{\tv-1}^{i}).
        \label{eq:weighted-prediction-approximation}
\end{align}
If the distribution $p(\vstate_{\tv} \mid \vstate_{\tv-1})$ can be simulated, \cref{eq:weighted-prediction-approximation} is also easily simulated by first sampling ancestor indices $a^{i}_{\tv}$ from a categorical distribution with probabilities proportional to $\{\w_{\tv-1}^{i}\}_{i=1}^{N}$, and then sampling $p(\vstate_{\tv} \mid \vstate^{a^{i}_{\tv}}_{\tv-1})$.
This provides an unweighted approximation of the predictive distribution $\what{p}(\vstate_{\tv} \mid \vobs_{\range{1}{\tv-1}}) = \sum_{i=1}^{N} \delta_{\vstate^{i}_{\tv}}(\vstate_{\tv})$.
Inserting this approximation in \cref{eq:update-step} yields a weighted approximation to the filtering distribution at time $\tv$ given by
\begin{align}
    \what{p}(\vstate_{\tv} \mid \vobs_{\range{1}{\tv}})
        &= \frac{\sum_{i=1}^{N} p(\vobs_{\tv} \mid \vstate_{\tv}^{i}) \delta_{\state_{\tv}^{i}}(\vstate_{\tv})}{\sum_{j=1}^{N} p(\vobs_{\tv} \mid \vstate_{\tv}^{j})}.
        \label{eq:weighted-update-approximation}
\end{align}
Comparing \cref{eq:weighted-prediction-approximation} and \cref{eq:weighted-update-approximation}, the normalized weights at time $\tv$ are 
\begin{align}
    \w_{\tv}^{i} = \frac{p(\vobs_{\tv} \mid \vstate_{\tv}^{i})}{\sum_{j=1}^{N} p(\vobs_{\tv} \mid \vstate_{\tv}^{j})}, \qquad i=1,\ldots,N.
    \label{eq:boostrap-wight-function}
\end{align}
The bootstrap particle filter is summarized in \cref{alg:bootstrap-particle-fitler}.
Note that this is a special case of \cref{alg:sequential-monte-carlo} with the unnormalized target distributions given by \eqref{eq:smc:joint-filter-target} and the proposal chosen as the state transition, $\q_{\tv}(\vstate_{\tv} \mid \vstate_{\tv-1}) = p(\vstate_{\tv} \mid \vstate_{\tv-1})$.

%
%
\setlength{\intextsep}{0pt}
\begin{algbackbox}
	\begin{algorithm}[H]
		\begin{algorithmic}
			\State Sample $\vstate_{1}^{i} \sim p(\vstate_{1})$ and compute the weights $\w_{\tv}^{i}$ using \cref{eq:boostrap-wight-function}.
			\For{$\tv=2,\dots,T$}
			\State (a) \textit{Resample:} Sample ancestor indices $\{a^{i}_{\tv}\}_{i=1}^{N}$ from a categorical \newline
			\hspace*{2.7em} distribution with probabilities $\{\w_{\tv-1}^{i}\}_{i=1}^{N}$. 
			\State (b) \textit{Propagate:} Simulate $\vstate_{\tv}^{i} \sim p(\vstate_{\tv} \mid \vstate_{\tv-1}^{a^{i}_{\tv}})$. \label{alg:bootstrap-particle-fitler:propagate}
			\State (c) \textit{Weight:} Compute weights $\w_{\tv}^{i}$ using \cref{eq:boostrap-wight-function}.
			\EndFor
		\end{algorithmic}
		\caption{Bootstrap particle filter (all steps are for $i=1,\ldots,N$)}
		\label{alg:bootstrap-particle-fitler}
	\end{algorithm}
\end{algbackbox}
\setlength{\intextsep}{12pt plus 2pt minus 2pt}

\end{tcolorbox}

\begin{tcolorbox}[title=Example: Twisted particle filter for the cascaded water tanks, breakable, enhanced, before upper={\parindent15pt\noindent}]
	One important and useful property of sequential Monte Carlo is that it provides unbiased estimates of the normalizing constant.
For models where the final unnormalized target distribution $\utarget_{T}(\vstate_{\range{1}{T}})$ equals the full probabilistic model $p(\vstate_{\range{1}{T}}, \vobs_{\range{1}{T}})$, the final estimate of the normalizing constant $\widehat{\Z}_{T}$ is an estimate of the likelihood $p(\vobs_{\range{1}{T}})$. 
Twisted particle filters give unbiased estimates for all twisting potentials. However, the choice of twisting potential affects properties of the likelihood estimate.
Using the optimal twisting potential yields an estimator with zero variance, since it produces the true likelihood with probability one.
For the cascaded water tanks model, an extended Kalman filter works well for estimating the latent states, but it does not give accurate estimates of the likelihood.
It is possible to combine the particle filter with an extended Kalman filter to approximate the optimal twisting potential. The resulting estimates of the likelihood are shown in \cref{fig:twisted_likelihood_estimates}.
The twisted filter is compared to a standard bootstrap filter, both use the transition distribution of the state-space model $p(\vstate_{\tv} \mid \vstate_{\tv-1})$ as proposal.
Both filters were run 100 times with the number of particles varying from 10 to 1\thinspace000.
Increasing the number of particles decreases the variance of the estimate.
The figure shows that the twisted particle filter allows for a reduction of the number of particles by an order of magnitude.
\begin{center}
	\includegraphics[width=0.77\columnwidth]{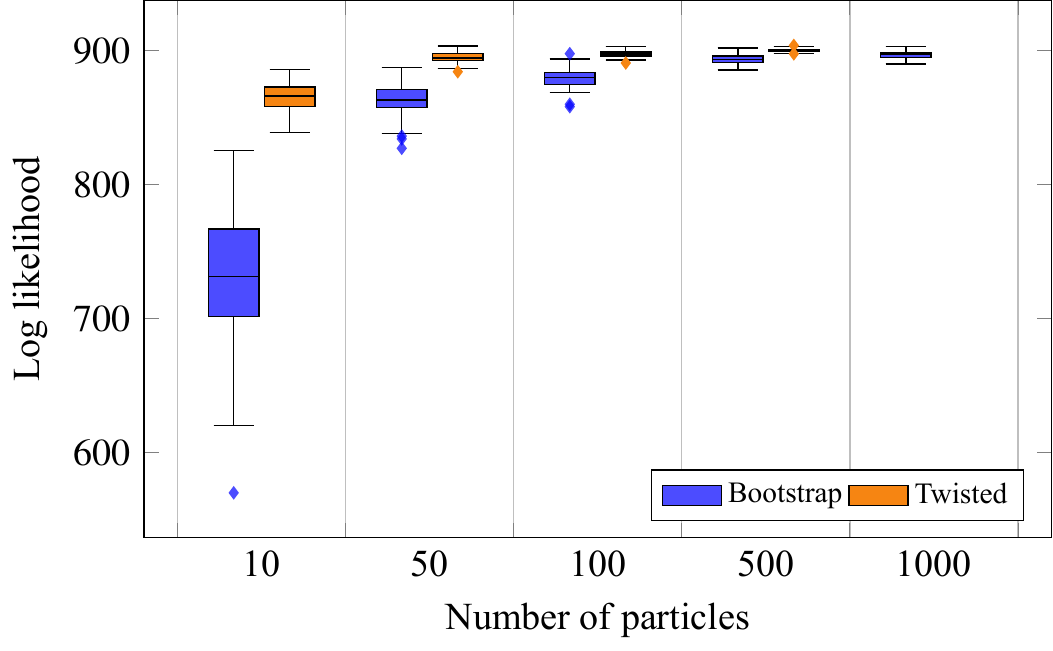}
\end{center}
\noindent \begin{minipage}{\textwidth}
	\captionof{figure}{
		Log likelihood estimates for the parameters produced by a bootstrap particle filter and a twisted particle filter where the twisting potential has been computed using an extended Kalman filter.
		Both filters produce unbiased estimates of the likelihood.
		The log-likelihood is negatively biased, since the logarithm is a concave function.
		Large variance gives both wider spread and more negative bias.
		The twisted particle filter requires approximately an order of magnitude fewer particles to produce estimates comparable to the bootstrap particle filter.
	}\label{fig:twisted_likelihood_estimates}
\end{minipage}       
\end{tcolorbox}

\section{Identification using marginalization}
This section briefly introduces the top row in \cref{fig:SIoverview}, that is, the marginalization strategies offered by direct gradient search and the MH algorithm when they are applied to state-space models.

\subsection{Direct gradient-based approach}
\label{sec:Direct}
In many cases, the ML optimization problem \cref{qe:maximum-likelihood} can be approached using standard gradient-based search.
The essential step of this approach involves iteratively updating the parameters via 
\begin{align}
    \vparam_{k+1} = \vparam_{k} + \alpha_k \vec{d}_k,
\end{align}
where $\vec{d}_k$ is a search direction, and $\alpha_k > 0$ is a so-called step-length that is determined so that the iterates converge.
For this strategy to be successful, it is enough to require that the search direction $\vec{d}_k$ is an ascent direction, which can be guaranteed for $\vec{d}_k = \nabla_{\param} \ln p(\vobs_{\range{1}{T}} \mid \vparam_k)$ \textemdash the gradient of the log-likelihood.

Ascent can be ensured more broadly by allowing any negative definite matrix $\mat{H}_k$ to scale the gradient, and importantly, this can lead to significant improvements in convergence rate.
A classical result along these lines reveals that the inverse Hessian matrix involving second-order curvature information is an ideal choice in this regard. 
Unfortunately, for the class of nonlinear state-space models considered in this article, it is not tractable to compute the log-likelihood, its gradient, or the associated Hessian matrix.
The essential reasons stem from the intractable integrals required in evaluating the log-likelihood, see \cref{eq:update-step,eq:prediction-step}. 

Given the success of SMC methods in this context, it is tantalizing to consider how they may be employed for approximating the gradient and Hessian as well.
The gradient and Hessian can be approximated using SMC methods with complexity that grows linearly in data length \cite{Nemeth2016}, as captured in \cref{alg:llGradHessApprox}.
This is just one of the methods that can be employed to compute the gradient and Hessian. There are many SMC-based alternatives, see \cite{Nemeth2016} and references therein for other approaches.
While the method outlined in \cref{alg:llGradHessApprox} provides estimates of the gradient and Hessian, these estimates are based on SMC methods and are therefore stochastic.
This raises a further issue in that standard gradient ascent methods are not intended to solve stochastic optimization problems. 

Fortunately, due to their importance, stochastic optimization problems are rather well studied.
The first known stochastic optimization algorithm was introduced in \cite{RobbinsM:1951}. It uses only first-order (gradient) information, motivating the name \textit{stochastic gradient (SG) method}.
Importantly, the majority of SG algorithms are not ascent methods. Rather, they are Markov chain methods, since their update rule defines a Markov chain.

Following the landmark article \cite{RobbinsM:1951}, many extensions have been developed within the statistics and automatic control communities.
Some notable works include convergence results~\cite{kiefer1952stochastic,ljung1977analysis,ljung1978strong}, online parameter estimation and system identification in~\cite{Ljung:1979,Ljung:1983}, adaptive control strategies~\cite{goodwin1981discrete}, and general books in the area~\cite{bertsekas1996neuro,spall2005introduction,ljung2012stochastic}.
The primary focus of current research activity is directed towards improving convergence rates.
Two important aspects that impact the convergence rate are:
\begin{itemize}
  \item
    Poor problem scaling, which can lead to slow convergence~\cite{BottouCN:2018}.
  \item
    Classical step-length formulas are conservative~\cite{AsiD20129siam,MoulinesB:2011}.
\end{itemize}
Regarding the first aspect, incorporating second-order information can greatly improve convergence, as highlighted in ~\cite{BottouCN:2018}.
Regarding the second aspect, the step-length can be determined using stochastic line-search procedures, modeled after the backtracking line-search with Armijo conditions~\cite{Armijo:1966,Wolfe:1969,Wolfe:1971}.
Recent work along these lines can be found in \cite{wills2019stochastic,WillsS:2020}.

In ``Example: Direct gradient-based search fr identification of cascaded water tank parameters'', the gradient-based search discussed above (with likelihood gradients and Hessians estimated using \cref{alg:llGradHessApprox}) is applied for learning the parameters of the coupled-tank model in ``Example: Cascaded water tanks''.
Different combinations of the parameters, corresponding to different physical interpretations of the model, are also evaluated.

\begin{algorithm}[htb]
    \caption{Compute log-likelihood gradient and Hessian estimate}
    \label{alg:llGradHessApprox}
    \begin{algorithmic}
        \State (a) Set $\vec{\alpha}_{0}^i = \0, \mat{\beta}_0^i = \0$, for $i=1,\ldots,N$.
        \State (b) Run \cref{alg:sequential-monte-carlo} to generate $\{\vstate_{\tv}^j, a_{\tv}^j\}_{j=1}^N$ and $\w_{\tv}^j$.
        \For{$\tv = 1, \dots, T$}
        	\State Compute
        	\begin{align}
        	\label{eq:pfNotes:30}
        	\vec{v}_{\tv} &= \sum_{j=1}^N \vec{\gamma}_{\tv}^j \w_{\tv}^j, \qquad 
        	\mat{B}_{\tv} = \vec{v}_{\tv} \vec{v}_{\tv}^\+ + \sum_{j=1}^N \Bigl [\mat{\phi}_{\tv}^j + \vec{\gamma}_{\tv}^j (\vec{\gamma}_{\tv}^j)^\+ \Bigr] \w_{\tv}^j,
        	\end{align}
        	\State where
        	\begin{subequations}
        		\begin{align}
        		\label{eq:6}
        		\vec{\gamma}_{\tv}^j &= \nabla_\param \log p(\vobs_{\tv} \mid \vstate_{\tv}^j, \vparam) + \nabla_\param \log p(\vstate_{\tv}^j \mid \vstate_{\tv-1}^{a_{\tv}^j}, \vparam) + \vec{\alpha}_{\tv-1}^{a_{\tv}^j}, \qquad 
        		\vec{\alpha}_{\tv}^j = \vec{\gamma}_{\tv}^j - \vec{v}_{\tv},\\
        		\mat{\phi}_{\tv}^j &= \nabla_{\param}^2 \log p(\vobs_{\tv} \mid \vstate_{\tv}^j, \vparam) + \nabla_{\param}^2 \log p(\vstate_{\tv}^j \mid \vstate_{\tv-1}^{a_{\tv}^j}, \vparam) + \mat{\beta}_{\tv-1}^{a_{\tv}^j}, \qquad \mat{\beta}_{\tv}^{j} = \mat{\phi}_{\tv}^j - \mat{B}_{\tv}.   
        		\end{align}
        	\end{subequations}
		\EndFor
        \State (c) Approximate $\nabla_{\param} \ln p(\vobs_{\range{1}{T}}\mid \vparam)$ and
        $\nabla_{\param}^2 \log p(\vobs_{\range{1}{T}} \mid \vparam)$ via
        \begin{align}
        \label{eq:pfNotes:32}
        \nabla_{\param} \ln p(\vobs_{\range{1}{T}} \mid \vparam) \approx \sum_{\tv=1}^{T} \vec{v}_{\tv}, \qquad 
        \nabla_{\param}^2 \ln p(\vobs_{\range{1}{T}} \mid \vparam) \approx \sum_{\tv=1}^{T} \mat{B}_{\tv}.
        \end{align}
	\end{algorithmic}
\end{algorithm}

\subsection{The Metropolis--Hastings algorithm}
\label{sec:MH}
The MH algorithm is perhaps the most well-known and most commonly used Markov chain Monte Carlo (MCMC) method, largely due to its simplicity.
A general, but brief, introduction to MCMC methods is available in ``Background: Markov chain Monte Carlo''.
For the MH algorithm, samples $\vstate$ from a target distribution $\target(\vstate)$ are generated by iteratively executing two steps.
First, a new candidate sample $\vstate^\prime$ is generated from a proposal distribution $\q(\vstate^\prime \mid \vstate(m))$, where $\vstate(m)$ is the most recently generated sample. 
The proposed sample $\vstate^\prime$ is then accepted or rejected according to the acceptance probability
\begin{equation} \label{eq:MHaccept}
	\alpha = \min \left(1,\frac{\target(\vstate^\prime)\q(\vstate(m) \mid \vstate^\prime)}{\target(\vstate(m)) \q(\vstate^\prime \mid \vstate(m))}\right). 
\end{equation}
If accepted, the proposed sample is kept and assigned as the next sample, $\vstate(m+1)=\vstate^\prime$. If rejected, the proposed sample is discarded and the most recently accepted sample is assigned as the next sample, $\vstate(m+1) = \vstate(m)$.
See \cite{casella2004} for a more detailed description of the algorithm.
An important observation is that the acceptance probability \cref{eq:MHaccept} contains a ratio of target distributions, implying that evaluation of the normalization constant is not required to compute the acceptance probability.
This is essential for practical applications, since the normalization constant is typically intractable.
Another observation is that the MH algorithm requires a proposal distribution, chosen by the user.
The proposal affects the convergence rate of the algorithm.
Ideally, the proposal should ensure fast exploration of the parameter space while maintaining a high acceptance probability.
In practice, however, a high acceptance probability typically aligns with a slow exploration and vice versa.
In particular, this is the case for high-dimensional problems.

For the system identification problems considered in this article, the target distribution of interest is the posterior distribution of the parameters, given in \cref{eq:Bayesian}.
The acceptance probability for this target is, by insertion in \cref{eq:MHaccept},
\begin{equation} \label{eq:targetMCMCBayes}
	\alpha = \min \left(1, \frac{p(\vobs_{\range{1}{T}} \mid \vparam^\prime) p(\vparam^\prime)}{p(\vobs_{\range{1}{T}} \mid \vparam(m))p(\vparam(m))}\frac{\q(\vparam(m)\mid \vparam^\prime)}{q(\vparam^\prime \mid \vparam(m))}\right),
\end{equation}
where the normalization constant $p(\vobs_{\range{1}{T}})$ in the target distribution \cref{eq:Bayesian} cancels.
The expression for the acceptance probability poses two difficulties: A proposal distribution must be chosen, and the likelihood $p(\vobs_{\range{1}{T}} \mid \vparam)$ must be evaluated.
The former (choosing a proposal) is in a sense simpler, since it is possible to run the algorithm for any proposal whose support contains the support of the target distribution.
The evaluation of the likelihood will, if intractable, prevent even running the algorithm.
An appealing idea is to replace the likelihood, when intractable, with an estimate.
This modification can be used to construct an algorithm that yields samples from the desired target distribution, provided the likelihood estimate is part of the generated Markov chain (likelihood estimate and parameters are accepted or rejected collectively), and the likelihood estimator is nonnegative and unbiased.
Such a likelihood estimator can be obtained from running SMC and evaluating \cref{eq:SMClikelihood}.
The resulting method belongs to the particle Markov chain Monte Carlo (PMCMC) family and is referred to as particle marginal Metropolis--Hastings (PMMH) \cite{AndrieuDH:2010}. 

The claim that any likelihood estimator that is unbiased and nonnegative can be used, deserves some motivation.
The most intuitive argument stems from deriving PMMH using a pseudo-marginal approach \cite{AndrieuR:2009}, where the target distribution is extended to include the likelihood estimate $\what{\Z}$.
A proposal distribution for the MH algorithm that yields the acceptance probability \cref{eq:targetMCMCBayes} with $p(\vobs_{\range{1}{T}} \mid \vparam)$ replaced with its estimated value, can be constructed by first sampling the parameters from some suitable proposal distribution, and then sampling $\what{\Z}$ from its distribution.
This approach is valid only if the extended target distribution is a valid probability distribution and the marginal distribution of the extended target with respect to the parameters is the true target distribution. These conditions are fulfilled only if the likelihood estimator is nonnegative and unbiased.
In \cite{SchonSML:2018} a tutorial-style derivation of this result is presented that is based on the original derivation in \cite{AndrieuR:2009}.

For the case considered here, where the likelihood estimate is obtained using SMC, an alternative proof that the resulting algorithm is valid is provided in \cite{AndrieuDH:2010}.
It too is based on extending the target distribution. However, the extension this time is by including all random variables that are generated in the SMC algorithm.
The PMMH sampler can then be obtained by designing a proposal distribution for this extended target distribution.

\begin{tcolorbox}[title=Example: Direct gradient-based search for identification of cascaded water tank parameters, breakable, enhanced, before upper={\parindent15pt\noindent}]
	In this section, the gradient-based search approach is demonstrated using the the cascaded water tank model described in \cref{eq:watertankSSM1,eq:watertankSSM2,eq:watertankSSM3} and the data from \cite{Schoukens2017}.
The data has $T = 1\thinspace024$ input--output measurements available for estimation and a further $T_v = 1\thinspace024$ input--output measurements available for validation purposes. 

Recall that the parameters are $\vparam=\{ k_1, k_2, k_3, k_4, k_5, k_6, \sigma_{\pnoise}^2, \sigma_{\onoise}^2 \}$, and for the purposes of illustration, they are initialized to 
\begin{align}
    \vparam_{1} = \{ 0.2, 0, 0.2, 0, 0.2, 0.2, 0.1, 0.1 \}.
\end{align}
While it is possible to also add the initial states as unknown parameters, they are simply chosen as $\vstate_{1} = [6, \obs_{1}]^\+$ here.
It is instructive to consider several other model structure choices to demonstrate their effect on the predictive performance.
Different model structures can be explored by eliminating selected terms in the model (by setting the accompanying coefficients to zero), and observing the predictive performance in each case.
It is important to notice the relative simplicity in changing the model structure, and its physical interpretation.
The latter is often difficult or impossible in more general black-box models.
In all cases considered, $N=50$ particles were used.

Consider the model where coefficients $k_2, k_4$, and $k_6$ are all zero.
This has the effect of removing the linear loss terms and the overflow event from the model.
Without the loss and overflow terms, the parameters are $\vparam = \{k_1, k_3, k_5, \sigma_{\pnoise}^2, \sigma_{\onoise}^2 \}$.
The gradient-search algorithm was allowed to run for $100$ iterations.
Using the obtained estimate in combination with the validation data input sequence, a simulated output, $\what{\obs}_{\tv}$, can be computed by iterating \cref{eq:watertankSSM1,eq:watertankSSM2,eq:watertankSSM3}.
The simulated output can be compared with the observed validation data output, $\obs_{\tv}$, to provide a measure of predictive performance by computing the rootmean-
square simulation error
\begin{align}
    \label{eq:rmsError}
    e_{\text{RMS}} = \sqrt{\frac{1}{T_v} \sum_{\tv=1}^{T_{v}} \left\|\obs_{\tv} - \what{\obs}_{\tv} \right\|^2}.
\end{align}
The predictive metric $e_{\text{RMS}}$ for the estimated parameters and their log-likelihood is captured in \cref{tab:gbs_results}. 

Inspired by the prospect that the model could benefit from including more terms, the linear loss coefficients $k_2$ and $k_4$ are added to the estimated parameter list.
After $100$ iterations, the predictive metric $e_{\text{RMS}}$ for the estimated parameters is captured in \cref{tab:gbs_results}.
While the log-likelihood value has increased, the predictive performance has degraded with these parameter additions.
This highlights the often-observed phenomena that improved training cost does not necessarily lead to improved predictive performance on validation data. 

The reverse situation is considered next, where $k_2$ and $k_4$ are set to zero, and the overflow event coefficient $k_6$ is allowed to be estimated.
\Cref{tab:gbs_results} reveals that the log-likelihood increased, and that the predictive performance also improved relative to both previous cases.
The implication is that the overflow event should be modeled.

Finally, the case that includes all parameters was tested, revealing the estimated values
\begin{align}
    \widehat{\vparam} = \{ 0&.0392, 0.0016, 0.0637, -0.0059, 0.0414, 0.2572, 0.0012, 0.0001 \}
\end{align}
The estimates of the two loss coefficients $k_2$ and $k_4$ are both an order of magnitude less than the estimated $k_1$ and $k_3$ coefficients. 
The log-likelihood and predictive metric are recorded in \cref{tab:gbs_results}, revealing that the log-likelihood has increased compared to the
\vspace{-3mm}
\begin{center}
	\begin{table}[H]
		\caption{
			Log likelihood and predictive performance for different parameter combinations for the cascaded water tank example.
			Higher log likelihood values indicate improved fit between the model predicted and the measured output.
			Lower $e_\text{RMS}$ values indicate reduced simulation error on validation data.
		}
		\label{tab:gbs_results}
		\centering
		\newcolumntype{g}{>{\columncolor{white}}c}
		\begin{tabular}{|g|g|g|}
			\hline
			Parameters & Log likelihood & $e_{\text{RMS}}$ \\
			\hline
			$k_{1,3,5}$, $\sigma_{\pnoise}^2$, $\sigma_{\onoise}^2$ & $1\thinspace388$ & $0.64$\\
			\hline
			$k_{1,2,3,4,5}$, $\sigma_{\pnoise}^2$, $\sigma_{\onoise}^2$ & $1\thinspace540$ & $0.90$\\
			\hline
			$k_{1,3,5,6}$, $\sigma_{\pnoise}^2$, $\sigma_{\onoise}^2$ & $1\thinspace810$ & $0.49$\\
			\hline
			$k_{\range{1}{6}}$, $\sigma_{\pnoise}^2$, $\sigma_{\onoise}^2$ & $1\thinspace979$ & $0.28$\\
			\hline
		\end{tabular}
	\end{table}
\end{center}
previously considered models, and the predictive performance on the validation dataset was best of all models tested.
A plot of the measured and simulated outputs is provided in \cref{fig:gbstanks}. 

\begin{figure}[H]
    \centering
    \includegraphics[width=0.7\textwidth]{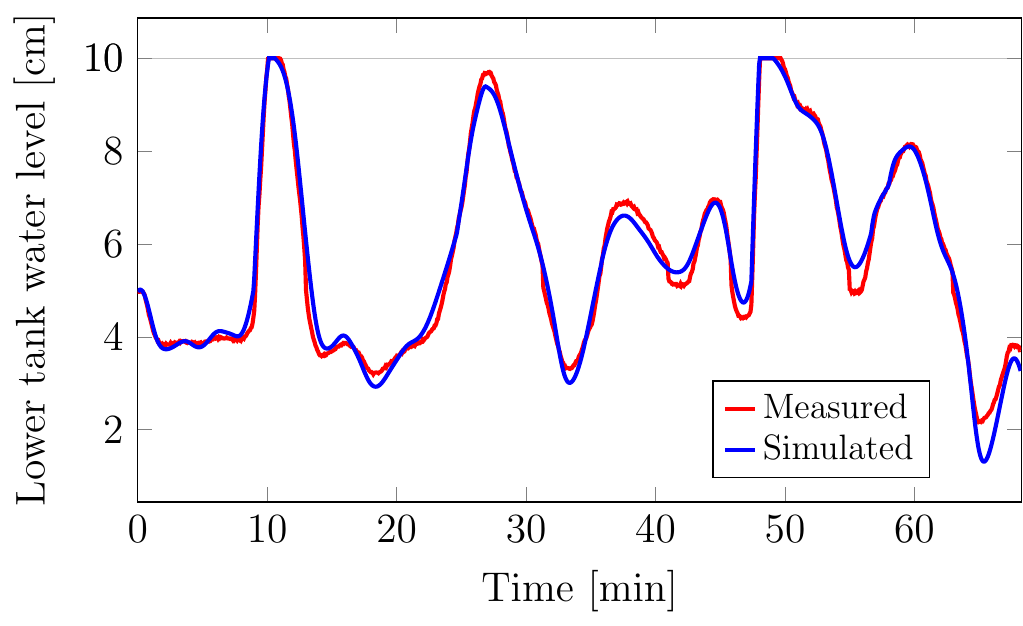}
	\caption{
        Measured and simulated outputs for direct gradient-based search on the cascaded water tank example.
        The validation data was not used for estimating the model, and the simulated output relies on the input signal and model only.\label{fig:gbstanks}}
\end{figure}
\end{tcolorbox}

\begin{tcolorbox}[title=Background: Markov chain Monte Carlo, breakable, enhanced, before upper={\parindent15pt\noindent}]
	Markov chain Monte Carlo (MCMC) methods are useful for evaluating expectation integrals that are otherwise challenging to compute.
In the context of Bayesian identification, MCMC methods can be employed to compute expectations relative to the posterior distribution $p(\vparam \mid \vobs_{\range{1}{T}})$ of the form 
\begin{align} 
    \label{eq:2}
    I = \int f(\vparam)\, p(\vparam \mid \vobs_{\range{1}{T}}) \;\rmd \vparam,
\end{align}
where the function $f(\cdot)$ may be quite general.
For example, $f$ could be the indicator function that $\vparam$ belongs to a given set. If $f(\vparam) = \vparam$, the above integral computes the conditional expected mean, which is also the minimum-mean-squared estimator~\cite{Kay:1993}.

Solving these types of integrals is generally intractable in closed-form, and classical quadrature methods are limited to low dimensions only.
An alternative computational approach relies on the law of large numbers to provide estimates of \cref{eq:2} via so-called Monte-Carlo integration, where $M$ samples $\{\vparam(m)\}_{m=1}^M$ from $p(\vparam \mid \vobs_{\range{1}{T}})$ are used to compute a sample average of $f(\cdot)$ according to
\begin{align}
    \label{eq:3}
    \what{I}_M = \frac{1}{M} \sum_{m=1}^M f(\vparam(m)), \qquad \vparam(m) \sim p\big(\vparam \mid \vobs_{\range{1}{T}}\big).
\end{align}
Provided that $p(\vparam \mid \vobs_{\range{1}{T}})$ and $f(\cdot)$ satisfy some weak assumptions \autocite{casella2004}, it can be shown that 
\begin{align}
    \label{eq:4}
    \what{I}_M \overset{\text{a.s.}}{\to} I \quad \text{as} \quad  M \to \infty,
\end{align}
where a.s. denotes almost sure convergence.
Importantly, the rate of convergence of $\what{I}_M$ to $I$ is maximized when the samples
$\vparam(m)$ are uncorrelated~\autocite{Tierney1994}.

Some natural questions arise concerning how to generate samples from $p(\vparam \mid \vobs_{\range{1}{T}})$ and how to sample so that the correlation is minimized.
A remarkably effective approach aimed at addressing these questions is to construct a Markov chain whose stationary distribution coincides with the target $p(\vparam \mid \vobs_{\range{1}{T}})$.
Perhaps as equally remarkable is that such a Markov chain can be constructed in a straightforward manner using the Metropolis--Hastings algorithm or other alternatives such as Gibbs sampling.
The combination of Monte Carlo integration with a Markov Chain for generating samples reveals the MCMC epithet.
\end{tcolorbox}

\section{Identification using data augmentation}
This section briefly introduces the bottom row in \cref{fig:SIoverview}, that is, the data augmentation strategies offered by expectation maximization and Gibbs sampling when they are applied to state-space models.

\subsection{Expectation maximization}
\label{sec:EM}
The EM approach \cite{DempsterLR:1977} is based on the idea that if (in addition to the output sequence $\vobs_{\range{1}{T}}$) the state sequence $\vstate_{\range{1}{T}}$ were known, then $\vparam$ could be estimated by solving the ML problem 
\begin{align}
    \widehat{\vparam} = \argmax_\param \ln p(\vobs_{\range{1}{T}},\vstate_{\range{1}{T}} \mid \vparam),
\end{align}
over the joint state--output data $p(\vobs_{\range{1}{T}},\vstate_{\range{1}{T}} \mid \vparam)$ in  \cref{eq:FullLikelihood}.
The above ML problem does \emph{not} involve any troubling integration terms, and in principle, this problem could be approached using standard optimization tools.
Furthermore, the solution $\widehat{\vparam}$ can be expressed in closed form in some cases, such as linear time-invariant state-space models \cite{gibson2005robust}.

Unfortunately, the state sequence $\vstate_{\range{1}{T}}$ is rarely available.
The EM approach combats this by replacing the joint log-likelihood $\ln p(\vobs_{\range{1}{T}},\vstate_{\range{1}{T}} \mid \vparam)$ with its expected value over the unobserved (or \emph{missing}) state sequence $\vstate_{\range{1}{T}}$, conditioned on the measured outputs $\vobs_{\range{1}{T}}$\textemdash the so-called \emph{expectation} or E-step.
It then aims to solve the surrogate \emph{maximization} problem (the M-step) 
\begin{align}
    \label{eq:EM:M}
    \vparam^\star = \arg \max_{\param} \int \ln p(\vobs_{\range{1}{T}},\vstate_{\range{1}{T}} \mid \vparam) p(\vstate_{\range{1}{T}} \mid \vobs_{\range{1}{T}}, \vparam^\prime) \;\rmd \vstate_{\range{1}{T}}
\end{align}
instead.
The conditional distribution $p(\vstate_{\range{1}{T}} \mid \vobs_{\range{1}{T}}, \vparam^\prime)$ relies on a different parameter $\vparam^\prime$, so that the variables being optimized over only appear in the joint log-likelihood term $\ln p(\vobs_{\range{1}{T}},\vstate_{\range{1}{T}} \mid \vparam)$.
The EM method progresses by solving for $\vparam^\star$, then updating $\vparam^\prime \leftarrow \vparam^\star$ and repeating.
This can be captured succinctly by indexing the parameters with iteration number $k$, so that starting with $\vparam_1$ repeatedly solves the problem 
\begin{align}
    \label{eq:EMequation} 
    \vparam_{k+1} = \argmax_{\param} \int \ln p(\vobs_{\range{1}{T}},\vstate_{\range{1}{T}} \mid \vparam) p(\vstate_{\range{1}{T}} \mid \vobs_{\range{1}{T}}, \vparam_k) \;\rmd \vstate_{\range{1}{T}},
\end{align}
for $k=1,2\dots$.
Importantly (and perhaps surprisingly), the sequence of $\vparam_k$'s is guaranteed not to decrease the log-likelihood \cite{DempsterLR:1977}, that is
\begin{align}
    \ln p(\vobs_{\range{1}{T}} \mid \vparam_{k+1}) \geq \ln p(\vobs_{\range{1}{T}} \mid \vparam_{k}),
\end{align}
which explains the primary mechanism for solving the maximum-likelihood problem \cref{qe:maximum-likelihood} using EM. 
Unfortunately, the expectation integral in \cref{eq:EMequation} is not generally tractable.
Nevertheless, in keeping with the theme of this article, this integral can be approximated using SMC methods that target smoothed distributions $p(\vstate_{\range{1}{T}} \mid \vobs_{\range{1}{T}}, \vparam_i)$.
In essence, the integral is approximated by a finite sum
\begin{align}
    \sum_{\npar=1}^{\Npar} \w^{\npar}(\vparam_k) \ln p(\vobs_{\range{1}{T}},\vstate_{\range{1}{T}}^i(\vparam_k)\,|\, \vparam) \approx \int \ln p(\vobs_{\range{1}{T}},\vstate_{\range{1}{T}}\,|\, \vparam) p(\vstate_{\range{1}{T}}\,|\, \vobs_{\range{1}{T}}, \vparam_k) \;\rmd \vstate_{\range{1}{T}}
\end{align}
where $\vstate_{\range{1}{T}}^{\npar}(\vparam_k)$ are particles and associated weights $\w^{\npar}(\vparam_k)$ from a particle smoother that depends on $\vparam_k$.
This approach has been explored by several authors \cite{CappeMR:2005, OlssonDCM:2008, SchonWN:2011}.
It gives rise to so-called particle expectation maximization (pEM) methods, where 
\begin{align}
    \label{eq:pEMequation} 
    \vparam_{k+1} = \argmax_\param \sum_{\npar=1}^{\Npar} \w^{\npar}(\vparam_k) \ln p(\vobs_{\range{1}{T}},\vstate_{\range{1}{T}}^{\npar}(\vparam_k) \mid \vparam)
\end{align}
is iterated.
Different particle smoothing algorithms can be used to address the underlying smoothing problem in the E-step of the EM algorithm, resulting in different variants of the pEM method.
When working with state-space models, one property that can be exploited is that the joint likelihood \cref{eq:FullLikelihood} factorizes over the time steps.
Therefore, even though \cref{eq:EM:M} seems to require the solution to a joint smoothing problem, it is enough to compute marginal smoothing estimates to implement the EM algorithm.
This enables, for example, \emph{fixed-lag smoothers} to be used, see \cite{OlssonDCM:2008} for details.
There are also extensions of pEM methods that couple the E-step and M-step to make more efficient use of the generated particles.
One such method is explained in ``Background: Particle stochastic approximation expectation maximization'', and demonstrated on the coupled-tank experiment in ``Example: Particle stochastic approximation expectation maximization for identification of cascaded water tank parameters''.

\subsection{Gibbs sampling}
\label{sec:Gibbs}
The Gibbs sampler \cite{GemanG:1984} is one of the most commonly used algorithms in the MCMC family.
Like all MCMC algorithms, it generates samples of some variable of interest $\vstate$ from its target distribution $\target(\vstate)$.
On a high level, the main idea behind the Gibbs sampler is to turn a potentially high-dimensional sampling problem into several subproblems of lower dimension that are, hopefully, simpler to sample from.
This is achieved by splitting the variable of interest $\vstate$ into smaller components, then updating these components by iteratively sampling them, one at a time, from their full conditional distributions while keeping all other components fixed.
A more mathematical description of the Gibbs sampler as well as a simple example is described in ``Example: Gibbs sampling''.

For the Bayesian formulation of the system identification problem considered in this article, the target distribution of interest is the parameter posterior, $p(\vparam \mid \vobs_{\range{1}{T}})$, given in \cref{eq:Bayesian}, with the parameters $\vparam$ being the variable of interest.
Targeting this distribution directly with a Gibbs sampler is difficult, since the intractable likelihood $p(\vobs_{\range{1}{T}}\mid\vparam)$ is required.
Instead, the Gibbs sampler targets the joint distribution $p(\vparam,\vstate_{\range{1}{T}} \mid \vobs_{\range{1}{T}})$, where the states have been introduced as auxiliary variables.
Gibbs sampling is therefore a method that most naturally belongs to the data augmentation strategy in \cref{fig:SIoverview}.
It is important to note that given that samples $\{\vparam(m),\vstate_{\range{1}{T}}(m)\}_{m=1}^M$ from the augmented target distribution $p(\vparam,\vstate_{\range{1}{T}} \mid \vobs_{\range{1}{T}})$ are available, the sequence of only the parameter samples $\{ \vparam(m) \}_{m=1}^M$ provides an approximation of the parameter posterior $p(\vparam \mid \vobs_{\range{1}{T}})$, since the parameter posterior is a marginal of the augmented target.
A Gibbs sampler that generates samples from the augmented target $p(\vparam,\vstate_{\range{1}{T}} \mid \vobs_{\range{1}{T}})$ can be designed to alternate between sampling a new state trajectory given the current parameters, $\vstate_{\range{1}{T}}(m+1)\sim p(\vstate_{\range{1}{T}} \mid \vparam(m),\vobs_{\range{1}{T}})$, and sampling new parameters given the current state trajectory, $\vparam(m+1)\sim p(\vparam \mid \vstate_{\range{1}{T}}(m+1),\vobs_{\range{1}{T}})$.
It is typically feasible to sample the parameters. However, sampling the state trajectory implies sampling from a high-dimensional, and in many cases intractable, distribution.
The distribution for generating new states can be recognized as the final target used in SMC for nonlinear state-space models.
However, simply applying \cref{alg:sequential-monte-carlo} to sample new states does not yield samples from the correct target distribution \cite{AndrieuDH:2010}.
In the following section on particle Gibbs (PG), a solution to this problem is discussed.

In contrast to other MCMC methods, like MH, no design of a proposal distribution is necessary in Gibbs sampling.
This is a clear advantage, especially when sampling from high-dimensional distributions where it can be difficult to find a suitable proposal.
On the other hand, Gibbs sampling performs poorly when there are strong dependencies between some of the components.

\subsubsection{Particle Gibbs}
The PG sampler generates samples from the joint posterior, $p(\vparam,\vstate_{\range{1}{T}} \mid \vobs_{\range{1}{T}})$, by alternately sampling new states conditioned on the current parameters and sampling new parameters conditioned on the current states.
To generate samples that have the correct target distribution, the sampling of new state trajectories must be done using a modified version of the SMC method in \cref{alg:sequential-monte-carlo}, referred to as conditional SMC \cite{AndrieuDH:2010}.
Conditional SMC is similar to ordinary SMC, but it takes a reference trajectory $\vstate^\prime_{\range{1}{T}}$ as input.
The reference trajectory is guaranteed to survive all resampling steps in the SMC algorithm, and at the final iteration, a new reference trajectory is generated according to the weights $\w_ T ^i$.
Intuitively, the reference trajectory can be thought of as a guide for the other trajectories that directs them to suitable parts of the state space.
A more detailed motivation and description of conditional SMC is provided in ``Background: Conditional sequential Monte Carlo'', and the complete PG sampler is outlined in \cref{alg:PG}.
In ``Example: Particle Gibbs for identification of dengue fever parameters'', PG is applied to learn the posterior distribution for the parameters of the model describing the spread of dengue fever (introduced in ``Example: Dengue fever'') for an outbreak of dengue fever in Micronesia \cite{Funk16}.
It also provides an illustration of how parameter priors for a Bayesian identification method can be designed based on attributes of the parameters and prior knowledge about the disease from previous studies.

\begin{algorithm}[htb] 
	\caption{The particle Gibbs sampler}
	\label{alg:PG}
	\begin{algorithmic}
		\State \textbf{Initialize:} Set $\vstate_{\range{1}{T}}(1)$ and $\vparam(1)$ arbitrarily.
		\For {$m=1 \dots M-1$}
			\State Run conditional SMC, inputs $\vstate_{\range{1}{T}}(m)$ and $\vparam(m)$, to draw $\vstate_{\range{1}{T}}(m+1)$. 
			\State Draw $\vparam(m+1)\sim p(\vparam \mid \vstate_{\range{1}{T}}(m+1),\vobs_{\range{1}{T}})$.
	    \EndFor
    \end{algorithmic}
\end{algorithm}

PG is part of a larger group of methods that combine MCMC methods with SMC, referred to as PMCMC methods, that were very briefly introduced in \cref{sec:MH}.
These methods are \textit{exact approximations} of the corresponding MCMC method in the sense that they return a sequence of samples with the desired target distribution, even though they use SMC to approximate some of the intermediate distributions \cite{AndrieuDH:2010}.
Any PMCMC method will, asymptotically as the number of MCMC iterations increases, yield samples from the correct target for any number of particles $N$ in the SMC component.
However, the performance usually improves with a larger number of particles.
For PG, a too low number of particles in the conditional SMC sampler leads to path degeneracy, that is, all samples share the same ancestry up to some time $\tv$.
This implies that all state trajectories collapse to the reference trajectory and, consequently, the state trajectory is rarely updated.
The result is a slow exploration of the space of state trajectories, and the PG sampler is said to ``mix poorly''.
The number of particles in the SMC component must be at least proportional to the number of time steps for good mixing, which is often prohibitive computation-wise in practice, since the complete algorithm then scales quadratically with time.
Even with proportionality between the number of particles and the time steps, the sampler can mix slowly for certain models~\cite{Lindsten2015}.

Fortunately, there are several extensions of the PG sampler that can significantly improve the mixing without significantly increasing the number of particles.
These extensions are typically based on trying to explore the state space around the reference trajectory more efficiently than is done in standard PG.
One example is particle Gibbs with backward sampling \cite{Whiteley2010,Lee2020} which, as the name suggests, uses backward sampling to draw a new reference trajectory.
Another example is particle Gibbs with ancestor sampling (PGAS) \cite{LindstenJS:2014}, which instead explores by sampling new ancestor indices for the reference trajectory.
PGAS is explained in more detail in ``Background: Particle Gibbs with ancestor sampling''.
A third possibility is to use blocking strategies, as suggested by \cite{Singh2017}.
This not only reduces the effect of path degeneracy, with improved theoretical and empirical stability as an effect, but also enables parallelization and adaptations of the algorithm for additional performance boosts.

\begin{tcolorbox}[title=Background: Particle stochastic approximation expectation maximization, breakable, enhanced, before upper={\parindent15pt\noindent}]
	The class of particle expectation maximization (pEM) methods are intuitively appealing, since they make use of particle smoothing algorithms in a natural way to address the intractable smoothing problem arising in the expectation maximization (EM) algorithm.
However, it has been realized that they sometimes make inefficient use of the particles that are generated.
At iteration $k$, $N$ particles are generated to approximately solve the E-step, followed by an update of the parameter from $\vparam_k$ to $\vparam_{k+1}$.
The generated particles are then discarded, and $N$ new particles must be generated at iteration $k+1$. However, the parameter update is often minor: $\vparam_{k+1} \approx \vparam_k$. In particular when the algorithm starts to converge to an optima.
Based on this observation, an appealing idea is to reuse the particles from iteration $k$ at iteration $k+1$ and so on.
One way to accomplish this is through stochastic approximation EM \cite{DelyonLM:1999} and its particle-based counterpart, particle stochastic approximation EM (PSAEM)~\cite{LindholmL:2019, Lindsten:2013}.

Without going into details, PSAEM tracks and updates smoothing estimates that are based on particles generated at \emph{all} previous iterations up to the current iteration $k$.
The ``old'' particles are gradually down-weighted based on a decreasing step-size sequence.
The method also makes use of a particle Markov chain Monte Carlo technique (reviewed in ``Particle Gibbs with ancestor sampling'') to create a dependence between particles generated at consecutive EM iterations.
When combined, this results in an algorithm that converges (as always, under some conditions) to a local maxima of the likelihood when $k \rightarrow \infty$, despite using a finite number of particles $N$ at each iteration. 
To borrow a phrase from \cite{LindholmL:2019}, the PSAEM algorithm ``entangles the convergence'' of the EM algorithm with the convergence of the underlying particle smoother.
The practical implication is that a convergent particle-based EM algorithm is obtained at a much lower computational cost than conventional EM methods. 
\end{tcolorbox}

\begin{tcolorbox}[title=Example: Particle stochastic approximation expectation maximization for identification of cascaded water tank parameters, breakable, enhanced, before upper={\parindent15pt\noindent}]
	In this section, the particle stochastic approximation expectation maximization (PSAEM) approach is demonstrated on the cascaded water tank model described in \cref{eq:watertankSSM1,eq:watertankSSM2,eq:watertankSSM3} using the data available from \cite{Schoukens2017}.
The setup is identical to that used in ``Example: Direct Gradient-Based Search for Identification
of Cascaded Water Tank Parameters''.
The PSAEM algorithm requires the user to select the number of particles $N$. For this example, $N=50$ was used.
The PSAEM method was allowed to run for $50$ iterations, yielding the final parameter estimates
\begin{align} \label{eq:PSAEMresult}
    \vparam_{50} = \{ 0&.0603, -0.0038, 0.0702, -0.0065, 0.0457, 0.2272, 0.0021, 0.0137 \}.
\end{align}
Similar to the direct approach, this estimate indicates that the two loss coefficients $k_2$ and $k_4$ are both slightly negative, and also an order of magnitude smaller than the estimated $k_1$ and $k_3$ coefficients.
For the estimated parameters, the root-mean-square simulation error \cref{eq:rmsError} was $e_{\text{RMS}} = 0.29$.
A plot of the measured and simulated outputs is provided in \cref{fig:psaemtanks}. 

\begin{figure}[H]
	\centering
    \includegraphics[width=0.7\textwidth]{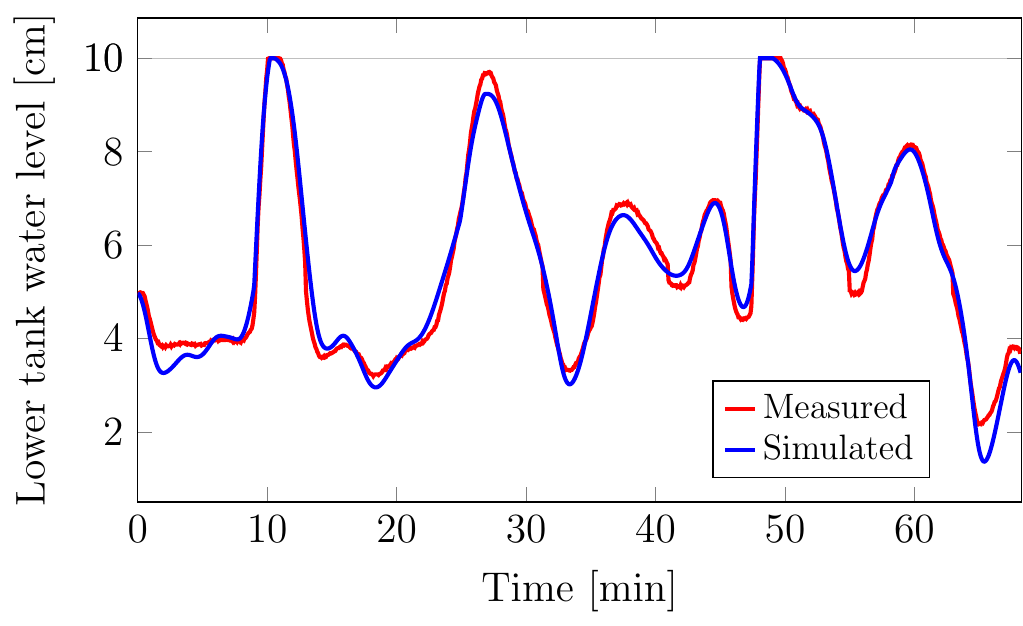}
	\caption{
        Measured and simulated outputs for particle stochastic approximation expectation maximization on the cascaded water tank example.
        The validation data was not used for estimating the model, and the simulated output relies on the input signal and model only.
        \label{fig:psaemtanks}
    }
\end{figure}
\end{tcolorbox}

\begin{tcolorbox}[title=Example: Gibbs sampling, breakable, enhanced, before upper={\parindent15pt\noindent}]
	Assume that samples of some random variable of interest $\vstate$ must be generated from a target distribution $\target(\vstate)$.
In addition, assume that the variable of interest can be split into $d$ components, $\vstate = (\vstate_1, \vstate_2,\dots,\vstate_ d)$.
Gibbs sampling can then be employed to iteratively generate samples from the target distribution by updating each component $\vstate_j$ in $\vstate$ separately according to its full conditional distribution $\target(\vstate_ j \mid \vstate_ 1, \dots ,\vstate_{j-1},\vstate_{j+1}, \dots ,\vstate_ d)$, while keeping all other components fixed.
Updating one component at a time in this way ensures that the sample is still distributed according to the desired target distribution $\target(\vstate)$ after each update.

To make this a bit more concrete, consider the case when $\vstate$ consists of three components $\vstate_1$, $\vstate_2$, and $\vstate_3$, and the target distribution of interest is $\target(\vstate) = \target(\vstate_1, \vstate_2, \vstate_3)$.
The Gibbs sampler is first initialized by assigning $\vstate(1) \sim p(\vstate(1))$, where $p$ is some initial distribution chosen by the user.
Gibbs sampling now proceeds by first sampling a new value for $\vstate_1$ conditioned on the current values for $\vstate_2$ and $\vstate_3$, that is, $\vstate_1(2) \sim \target \big(\vstate_1 \mid \vstate_2 (1), \vstate_3 (1) \big)$.
Next, given the current values for $\vstate_1$ and $\vstate_3$, a new value for the second component, $\vstate_2(2)$, is generated from $\target\big(\vstate_2 \mid \vstate_1 (2), \vstate_3 (1) \big)$.
Finally, a new sample $\vstate_3(2)$ is generated, given the current values of $\vstate_1$ and $\vstate_2$, from $\target\big(\vstate_3 \mid \vstate_1 (2), \vstate_2 (2) \big)$.
The new sample is $\vstate(2) = \big(\vstate_1(2),\vstate_2(2),\vstate_3(2)\big)$.
Gibbs sampling then proceeds by repeating these three steps to generate more samples.
Eventually, as the number of samples increases, the samples will be distributed according to the target distribution.
The procedure is summarized in \cref{alg:Gibbs}.

\setlength{\intextsep}{0pt}
\begin{algbackbox}
	\begin{algorithm}[H] 
		\caption{The Gibbs sampler for the target $\target(x)=\target(\vstate_1,\vstate_2,\vstate_3)$}
		\label{alg:Gibbs}
		\begin{algorithmic}
			\State \textbf{Initialize:} For $\vstate=(\vstate_1, \vstate_2, \vstate_3)$, set $\vstate(1) \sim p\big(\vstate(1)\big)$
			\For {$m=1 \dots M-1$}
				\State $\vstate_1 (m+1) \sim \target\big(\vstate_1 \mid \vstate_2 (m), \vstate_3 (m)\big)$,
				\State $\vstate_2 (m+1) \sim \target\big(\vstate_2 \mid \vstate_1 (m+1), \vstate_3 (m)\big)$,
				\State $\vstate_3 (m+1) \sim \target\big(\vstate_3 \mid \vstate_1 (m+1), \vstate_2 (m+1)\big)$.
			\EndFor
		\end{algorithmic}
	\end{algorithm}
\end{algbackbox}
\setlength{\intextsep}{12pt plus 2pt minus 2pt}	
\end{tcolorbox}

\begin{tcolorbox}[title=Background: Conditional sequential Monte Carlo, breakable, enhanced, before upper={\parindent15pt\noindent}]
	Conditional sequential Monte Carlo (SMC) is a version of the method presented in \cref{alg:sequential-monte-carlo} that takes a reference trajectory $\vstate^\prime_{\range{1}{T}}$ as input and generates a new reference trajectory as output.
The name conditional SMC stems from the fact that the method generates samples like SMC, but it conditions on the reference trajectory surviving all resampling steps.
In each step of conditional SMC, $N-1$ ancestor indices are generated according to the weights of each particle in the same way as in SMC.
The last ancestor index, however, is always set to be the index of the reference trajectory.
In the propagation step, $N-1$ new states are simulated according to the proposal distribution.
The last state is set deterministically to the reference trajectory.
Finally, the weights are updated using \cref{eq:weight-function} and normalized.
After the final iteration, a new reference trajectory is generated based on the weights $\w_{T}^i$.
Conditional SMC is summarized in \cref{alg:cSMC}.

\setlength{\intextsep}{0pt}
\begin{algbackbox}
	\begin{algorithm}[H] 
		\caption{Conditional SMC (all steps for $i=1,\dots,N$)}
		\label{alg:cSMC}
		\begin{algorithmic}
			\State \textbf{Input:} Reference trajectory $\vstate_{\range{1}{T}}^\prime$, parameters $\vparam$, observations $\vobs_{\range{1}{T}}$.
			\State Draw $\vstate_{1}^{i} \sim \q_{1}(\vstate_{1})$ independently, set $\vstate_{1}^{N} = \vstate_{1}^\prime$.
			\State Compute weights $\uw_{1}^i = \utarget_1(\vstate_{1}^{i})/\q_1(\vstate_{1}^{i})$ and normalize $\w_1^ i = \uw_1^i / \sum_{j=1}^N \uw_1^j$.
			\For {$\tv=2 \dots T$}
				\State (a) \textit{Resample:} Draw ancestor indices $\{a_{\tv}^{i}\}_{i=1}^{N-1}$ with probabilities $\{\w_{\tv-1}^ i\}_{i=1}^N$ \newline \hspace*{2.7em} and set $a_{\tv}^N = N$. Set $\w_ {\tv-1}^i = 1/N$.
				\State (b) \textit{Propagate:} Simulate $\vstate_{\tv}^{i} \sim \q(\vstate_{\tv} \mid \vstate_{\tv-1}^{a_{\tv}^i})$ and set $\vstate_{\tv}^{N} = \vstate_{\tv}^\prime$. Set \newline \hspace*{2.7em} $\vstate_{\range{1}{\tv}}^{i}=\{ \vstate_{\range{1}{\tv-1}}^{a_{\tv}^i}, \vstate_{\tv}^{i} \}$.
				\State (c) \textit{Weight:} Set $\uw_{\tv}^{i} = \w_{\tv}(\vstate_{\range{1}{\tv}}^{i})$ and normalize $w_ {\tv}^i = \uw_{\tv}^i \sum_{j=1}^N \uw_{\tv}^j$.
				\EndFor
			\textbf{Output:} Draw $j$ with probability $\{\w_{T}^{i}\}_{i=1}^{N}$, output new reference trajectory $\vstate_{\range{1}{T}}^\prime = \vstate_{\range{1}{T}}^{{j}}$.
		\end{algorithmic}
	\end{algorithm}
\end{algbackbox}
\setlength{\intextsep}{12pt plus 2pt minus 2pt}	
\end{tcolorbox} 

\begin{tcolorbox}[title=Example: Particle Gibbs for identification of dengue fever parameters, breakable, enhanced, before upper={\parindent15pt\noindent}]
	In this sidebar, particle Gibbs (PG) is used to identify the posterior distribution of the parameters in the model describing the spread of dengue fever introduced in ``Example: Dengue fever'', using the dataset from the dengue outbreak on Yap \cite{Funk16}.
In PG, the Bayesian view on the parameters is adopted, which requires the specification of prior distributions for the parameters. 
The choices of parameter priors and initial values are discussed first, followed by the inference results for the dengue dataset from Yap.

\subsubsection*{Choosing a parameter prior}
The unknown model parameters are $\vparam = \{ \lambda^h, \delta^h, \gamma^h, \lambda^m, \delta^m, \gamma^m, \rho \}$.
All parameters are probabilities, which implies that they can only take values in the range $[0,1]$.
The beta distribution $\text{Beta}(\alpha,\beta)$, with shape parameter $\alpha$ and scale parameter $\beta$, is a common choice for modeling such a random variable.
To fully specify the beta prior, values for the distribution's parameters $\alpha$ and $\beta$ must be selected.
The prior should reflect the beliefs about the parameters before seeing the data and can, for instance, be based on results from previous studies of the disease.

The incubation and infection times for humans, as well as the incubation time for mosquitoes are well-studied and can be used to specify priors for the infection and recovery probabilities $\delta$ and $\gamma$.
In \cite{Funk16}, the rate parameters in \cref{eq:DengueStateUpdate} are specified through Gaussian priors on the incubation and infection times, which are related to the corresponding transition rates as rate=1/time.
Mode-matching can be used to incorporate the information from these Gaussian priors into the beta priors.
The mode for a Gaussian distribution is simply its mean value.
For the beta distribution, the mode is $m=\frac{\alpha-1}{\alpha+\beta-2}$.
Assuming that the mean transition time is $\mu_0$, the mean transition rate is $1/\mu_0$, and mode-matching results in the relation 
\begin{equation}\label{eq:mode-match}
    \frac{\alpha-1}{\alpha+\beta-2} = \frac{1}{\mu_0}.
\end{equation}
One choice fulfilling \eqref{eq:mode-match} is $\alpha=1+\frac{2}{\mu0}$ and $\beta=3-\frac{2}{\mu0}$.
From \cite{Funk16}, $\mu_0=4.4$ for the human incubation time, $\mu_0=4.5$ for the human infectious time, and $\mu_0=6.5$ for the mosquito incubation time.
Hence, the beta prior is $\delta^h\sim\text{Beta}(1+\frac{2}{4.4}, 3-\frac{2}{4.4})$ for the human infection probability, $\gamma^h\sim\text{Beta}(1+\frac{2}{4.5}, 3-\frac{2}{4.5})$ for the human recovery probability, and $\delta^m\sim\text{Beta}(1+\frac{2}{6.5}, 3-\frac{2}{6.5})$ for the mosquito infection probability.

To understand why the rates in \cite{Funk16} can be directly related to the corresponding probabilities in the probabilistic formulation, \cref{eq:LawProb,eq:DengueStateUpdate}, note that the binomial transitions imply geometrically distributed transition times.
A geometric distribution with mean transition time $z$ has a mean transition probability $1/z$. Hence, the transition probabilities, for which a prior is designed, are related to the mean transition time in exactly the same way as the transition rates in \cite{Funk16} are.
This motivates using the mean transition rate as a proxy for centering the beta prior on the transition probabilities.

The transmission times are unknown for both humans and mosquitoes.
The prior on the transmission probabilities $\lambda^h$ and $\lambda^m$ is therefore chosen to have parameters $\alpha=\beta=1$, which yields a uniform distribution between 0 and 1 to reflect this uncertainty.
The reporting probability is also unknown apriori.
Following the same reasoning as for the transmission probabilities, the prior on the reporting probability is chosen to have $\alpha=\beta=1$.
Finally, it was noted in the model description that the mosquitoes never recover from the disease.
Based on this, the prior for the recover probability for mosquitoes is set to zero.
In conclusion, the parameter priors are
\begin{equation}
    \begin{aligned}
        \lambda^h &\sim \text{Beta}(1,1),                              & \lambda^m &\sim \text{Beta}(1,1),                    \\
        \delta^h  &\sim \text{Beta}(1+\frac{2}{4.4},3-\frac{2}{4.4}),  & \delta^m  &\sim \text{Beta}(1+\frac{2}{6.5},3-\frac{2}{6.5}),  \\
        \gamma^h  &\sim \text{Beta}(1+\frac{2}{4.5},3-\frac{2}{4.5}),  & \gamma^m  &=0,\\
        \rho      &\sim \text{Beta}(1,1).
    \end{aligned}
    \label{eq:denguePrior}
\end{equation}
The parameter priors are visualized in \cref{fig:betaPriors}.
\begin{figure}[H]
	\centering
	\includegraphics[width=0.7\textwidth]{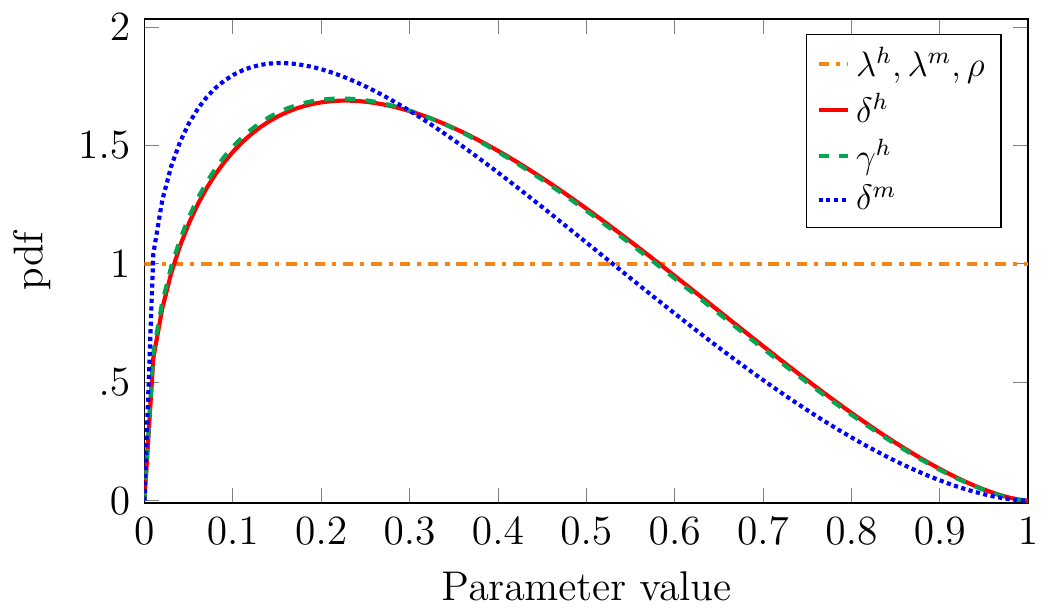}
	\caption{
        The different beta priors used for the parameters of the model describing the spread of dengue fever.
        Note that the priors for the parameters $\lambda^h$, $\lambda^m$, and $\rho$ are uniform, which reflects the uncertainty about these parameter values.
        The priors for the other parameters\textemdash $\delta^h$, $\gamma^h$, $\delta^m$\textemdash are skewed and unimodal, which reflects that they are informed by the prior knowledge about incubation and infection times.
    }
	\label{fig:betaPriors}
\end{figure}

\subsubsection*{Choosing initial values for the states}
The initial values are set to
\begin{equation}
    \begin{aligned}
    	E^{h}_{1}  &\sim \text{Poisson}(5),	 & E^{m}_{1}&=0,  \\
    	I^{h}_{1}-1&\sim \text{Poisson}(5),  & I^{m}_{1}&=0,  \\
    	R^{h}_{1}  &\sim \text{Uniform}(0,7370-E^{h}_{1}-I^{h}_{1}), 	 & R^{m}_{1}&=0,  \\
    	S^{h}_{1}  &=7370-E^{h}_{1}- I^{h}_{1}- R^{h}_{1},	 & S^{m}_{1}&=10^{u}\cdot 7370, 
    \end{aligned}
    \label{eq:initials}
\end{equation}
with $u\sim\mathcal{U}(-1,2)$.
The Poisson distribution for the initial number of exposed and infectious humans reflects the belief that, on average, five people are expected to be exposed and infectious initially, and the infections occur independent of each other.
The $-1$ in the expression for the initial number of infectious ensures that there is always at least one infectious individual.
There have been previous outbreaks of dengue fever on Yap, so the number of immune individuals in the human population is not known apriori.
Therefore, the initial number of recovered is sampled from a uniform distribution.
The mosquitoes are assumed to not recover from the disease, which motivates setting the number of recovered to zero initially.
Additionally, the lifespan of mosquitoes is short in comparison with that for humans. Thus, it is reasonable to assume that initially the spread of the disease originates from the human population.
Hence, the number of exposed and infectious mosquitoes are initially set to zero.
Finally, it is unknown how large the mosquito population is.
Consequently, it is initialized to be in a large range, allowing for it to be both smaller and much larger than the human population.

\subsubsection*{Inferring the parameters from data using particle Gibbs}
A PG sampler was run for $M=10\thinspace000$ iterations with $N=1\thinspace024$ particles in the SMC component.
The parameter priors and initial values were selected according to \cref{eq:denguePrior} and \cref{eq:initials}.
The simulation was repeated four times.
A histogram for the reporting probability $\rho$ (average of four runs) is shown in \cref{fig:DengueHist}.
It can be seen that the reporting probability is likely to be somewhere between $0.2$ and $0.5$, implying that a large proportion of the dengue cases are never reported to a health center.
It is interesting to note that the observed data from the outbreak has led to an update of the prior beliefs about the distribution of the reporting rate from a uniform prior distribution to the posterior in \cref{fig:DengueHist}.
Similar histograms can be generated for the other parameters of the model.

\begin{figure}[H]
	\centering
	\includegraphics[width=0.7\textwidth]{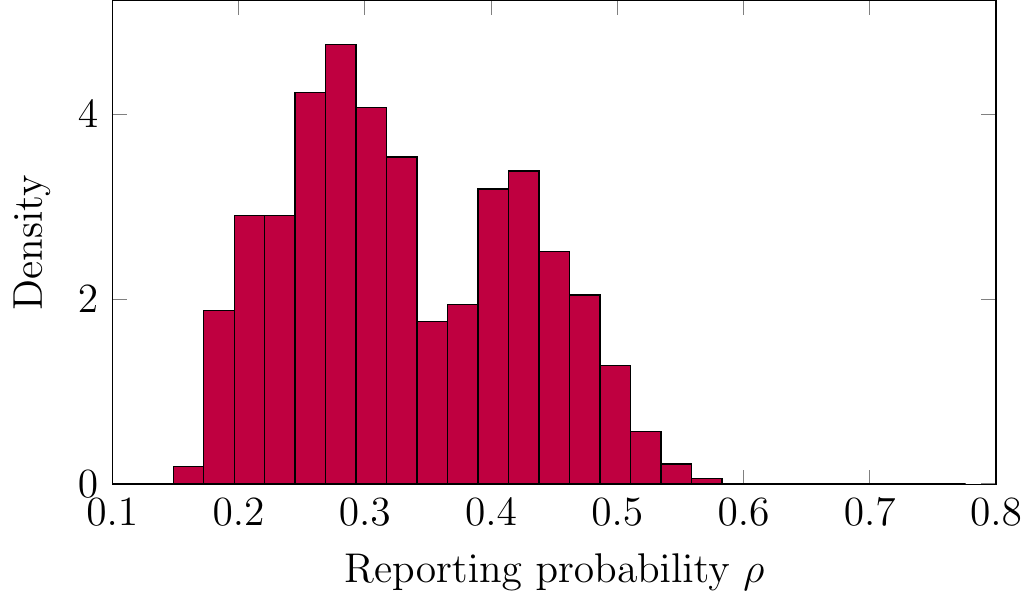}
	\caption{
        Histogram for the reporting probability $\rho$ in the dengue model.
        A particle Gibbs sampler (\cref{alg:PG}) was run for $M=10\thinspace000$ iterations using $N=1\thinspace024$ particles in the sequential Monte Carlo component.
        The parameter prior is a uniform distribution between $0$ and $1$, and the initial values are chosen according to \cref{eq:initials}.
        The histogram shows samples obtained over $4$ such runs.
        It can be seen that the reporting probability is likely to be somewhere between $0.2$ and $0.5$, implying that a large proportion of the dengue cases are never reported to a health center.
    }
	\label{fig:DengueHist}
\end{figure}
\end{tcolorbox}

\begin{tcolorbox}[title=Background: Particle Gibbs with ancestor sampling, breakable, enhanced, before upper={\parindent15pt\noindent}]
	Particle Gibbs with ancestor sampling (PGAS) is an extension of particle Gibbs (PG) that improves the mixing of the sampler by selecting a new ancestor for the reference trajectory at each point in time, instead of setting the ancestor deterministically (as is done in PG).
The ancestor of the reference trajectory $a_{\tv}^N$ is sampled according to the ancestor weights
\begin{equation} \label{eq:PGASweight}
	\wbar{\w}_{\tv-1 \mid T}^{i}\propto \w_{\tv-1}^{i} \frac{\utarget_{T}([\vstate_{\range{1}{\tv-1}}^{i},\vstate_{\range{\tv}{T}}^\prime] \mid \vparam)}{\utarget_{\tv-1}(\vstate_{\range{1}{\tv-1}}^{i} \mid \vparam)},
\end{equation} 
where $\w_{\tv-1}^{i}$ is the weight of the possible ancestor trajectory $i$, and $[\vstate_{\range{1}{\tv-1}}^{i},\vstate_{\range{\tv}{T}}^\prime]$ is a concatenation of the possible ancestor trajectory and the remaining part of the reference trajectory at time $\tv$.
Apart from the sampling of ancestor weights, the conditional sequential Monte Carlo (SMC) part of the PGAS algorithm, given in \cref{alg:cSMCPGAS}, is identical to \cref{alg:cSMC}.

In spite of being such a small change algorithm-wise, the sampling of new ancestor indices has a significant impact on the mixing of the sampler.
\Cref{fig:ACF_PGvsPGAS} shows an example of the improved performance over standard PG offered by PGAS in terms of the autocorrelation of the generated samples for the linear-Gaussian state-space model in \cref{eq:LGexample} with no input signal and unknown noise variances $Q$ and $R$. The prior on the noise variances is chosen to be an inverse-gamma distribution, which is a standard choice. Details on the experiments are in the figure caption.

\setlength{\intextsep}{0pt}
\begin{algbackbox}
	\begin{algorithm}[H]
		\caption{Conditional SMC with ancestor sampling (all steps for $i=1,\dots,N$)}
		\label{alg:cSMCPGAS}
		\begin{algorithmic}
			\State \textbf{Input:} Reference trajectory $\vstate_{\range{1}{T}}^{\prime}$, parameters $\vparam$, observations $\vobs_{\range{1}{T}}$.
			\State Draw $\vstate_{1}^{i} \sim \q_{1}(\vstate_{1})$ independently, set $\vstate_{1}^{N}=\vstate_{1}^{\prime}$.
			\State Compute weights $\uw_{1}^i = \utarget_{1}(\vstate_{1}^{i}) / \q_1(\vstate_{1}^{i})$ and normalize $\w_1^ i = \uw_1^ i / \sum_{j=1}^N \uw_1^j$. 
			\For {$\tv=2 \dots T$}
			\State (a) \textit{Resample:} Draw ancestor indices $\{a_{\tv}^{i}\}_{i=1}^{N-1}$ with probabilities $\{\w_{\tv-1}^i\}_{i=1}^N$ \newline \hspace*{2.7em} and draw $a_{\tv}^N$ with probabilities $\wbar{\w}_{\tv-1 \mid T}^i$ in \cref{eq:PGASweight}. Set $\w_ {\tv-1}^i = 1/N$. 
			\State (b) \textit{Propagate:} Simulate $\vstate_{\tv}^{i} \sim \q(\vstate_{\tv} \mid \vstate_{\tv-1}^{a_{\tv}^i})$ and set $\vstate_{\tv}^{N} = \vstate_{\tv}^\prime$. Set \newline \hspace*{2.7em} $\vstate_{\range{1}{\tv}}^{i}=\{ \vstate_{\range{1}{\tv-1}}^{a_{\tv}^i}, \vstate_{\tv}^{i} \}$. 
			\State (c) \textit{Weight:} Set $\uw_{\tv}^{i} = \w_{\tv}(\vstate_{\range{1}{\tv}}^{i})$ and normalize $\w_{\tv}^i = \uw_{\tv}^i / \sum_{j=1}^N \uw_{\tv}^j$. 
			\EndFor
			\State \textbf{Output:} Draw $j$ with probability $\{\w_{T}^{i}\}_{i=1}^{N}$, output new reference trajectory $\vstate_{\range{1}{T}}^\prime = \vstate_{\range{1}{T}}^{{j}}$.
		\end{algorithmic}
	\end{algorithm}
\end{algbackbox}
\setlength{\intextsep}{12pt plus 2pt minus 2pt}	

The improved performance is, perhaps surprisingly, not due to avoidance of path degeneracy.
In fact, path degeneracy also occurs for PGAS.
However, in PGAS the reference trajectory at each point in time is assigned different ancestors, which causes a collapse of the state trajectory to a trajectory that (with high probability) is different from the current reference trajectory.
Because of this small change, PGAS can perform well with a much lower number of particles $N$ in the SMC component than what is required for the corresponding PG sampler. 

\begin{figure}[H]
	\centering
	\includegraphics[width=.7\textwidth]{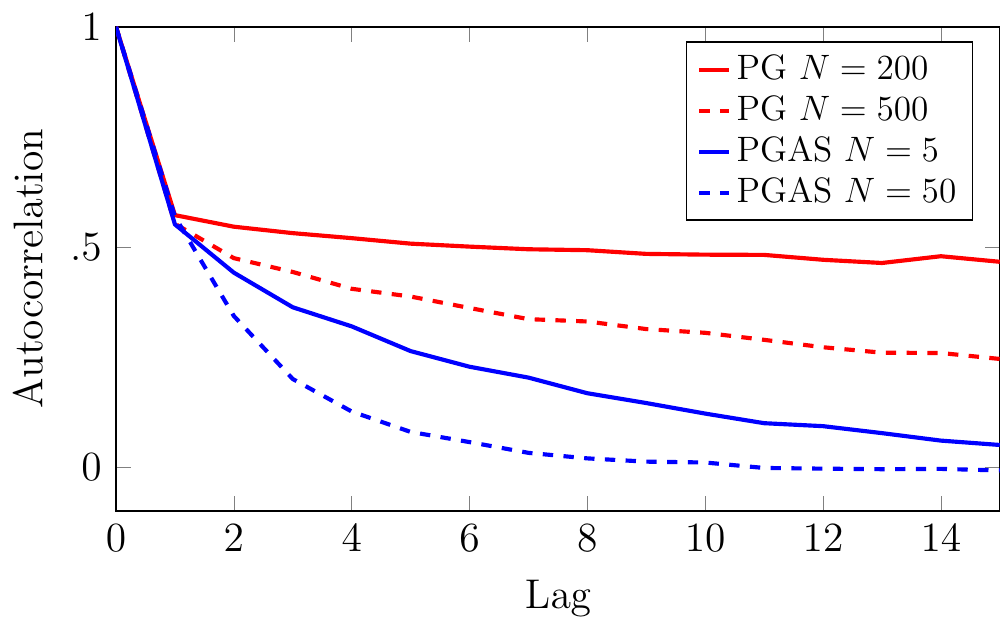}
	\caption{
		The autocorrelation function of the generated samples for particle Gibbs (PG) (red) and particle Gibbs with ancestor sampling (PGAS) (blue) for some different number of particles $N$ in the sequential Monte Carlo component.
		The figure shows the autocorrelation function of the noise covariance $Q$ in the linear-Gaussian state-space model \cref{eq:LG-transition,eq:LG-observation} for the one-dimensional case with no input $\inp$ and all parameters known except for the noise variances $Q$ and $R$.
		An inverse-gamma prior was used for both noise covariances, and there were $T = 150$ observations.
		Both samplers were allowed to run for $M=10\thinspace000$ iterations for two different choices of the number of particles $N$ for each sampler.
		The area under the autocorrelation function is related to the Monte Carlo variance of the generated samples.
		A smaller area corresponds to a smaller Monte Carlo variance and better mixing of the sampler.
		Thus, the figure indicates that PGAS performs better than standard PG, despite using a lower number of particles.
	}
	\label{fig:ACF_PGvsPGAS}
\end{figure}

\end{tcolorbox}

\section{Probabilistic programming}
To solve the nonlinear identification problems discussed in this article, both the physically motivated state-space model and the SMC-based inference method must be implemented in some programming language.
Implementing these inference methods in a standard programming language\textemdash like Matlab, C, or Python\textemdash can be a tedious and error-prone process.
Moreover, the end result is in many cases a model-specific implementation of the inference method that cannot straightforwardly be reused for another model.
Probabilistic programming is a recent tool that has been tailored to facilitate the implementation process by providing the user with a framework for encoding probabilistic models and performing inference in them in an automated way.
In a probabilistic program, the model specification and the inference algorithm are separated, which enables the user to focus on the modeling\textemdash the inference algorithm is already implemented.

\subsection{Representing the model as a program}
When specifying a model of a process, some type of modeling ``language'' is used to communicate the assumptions made about the behavior of that process.
In this article, models have so far mainly been described using the language of mathematics.
\Cref{fig:SSM} illustrates another modeling language\textemdash the graphical model.
Such a model explicitly shows the conditional dependence between the variables.
A probabilistic program is yet another modeling language, where the model is encoded as a computer program written in a Turing-complete probabilistic programming language (PPL).
Writing the model as a program allows for using programmatic constructs, such as stochastic branching and recursion, in the model formulation.
Constructs like these make probabilistic programs a much more expressive class of models than, for example, graphical models.
Consider \cref{fig:PPstochB}, which depicts a probabilistic program that incorporates stochastic branching.
Despite its seemingly simple form, this model cannot be formulated as a graphical model using any conventional representation.
\begin{figure}[htb]
	\centering
	\begin{minipage}{.14\textwidth}
		\hspace{1pt}
	\end{minipage}%
	\begin{minipage}{.5\textwidth}
	    \begin{algorithmic}[1]
			\State \texttt{x}$\sim$ \texttt{Gaussian(1,2);}
			\If {\texttt{(x>2)}}
			\State \texttt{y}$\sim$ \texttt{Student(x,3,2);}
			\Else \State \texttt{y}$\sim$ \texttt{Gaussian(x,2);}
			\EndIf
		\end{algorithmic}
	\end{minipage}%
	\caption{
		A probabilistic program with stochastic branching.
		The program first samples $x$ from a Gaussian distribution with mean value 1 and variance 2.
		Sampling is indicated by the symbol $\sim$.
		Depending on the sampled value for $x$, the variable $y$ is either sampled from a Student's t distribution or a Gaussian distribution.
	}
	\label{fig:PPstochB}
\end{figure}

The PPLs used to encode probabilistic programs are often based on already existing programming languages.
However, they differ from standard programming languages in two aspects: They have special constructs for conditional distributions, and they have an inference engine that is in control of the execution of the program.
To make the distinction between a standard computer program and a probabilistic program clearer, consider a standard computer program.
It is provided with some inputs and is then executed deterministically, step by step, to produce some output.
A probabilistic program instead takes observations as inputs and is executed deterministically only until it reaches a so-called \textit{checkpoint}, where the execution is paused.
At the checkpoint, the inference engine takes the current state of the program, manipulates it as specified in the program, returns the updated state and resumes deterministic execution of the program.
The checkpoints, where all randomness is introduced into the program, can be of different types.
A typical setup is to have at least two basic checkpoints: \textit{sample}, which creates a random variable and can trigger sampling of that variable; and \textit{observe}, which triggers conditioning on observed data \cite{Anglican2016}.
\Cref{fig:PPlgModel} shows a probabilistic program of a linear-Gaussian state-space model that incorporates both of these checkpoints.
A more complex example is provided in ``Example: Dengue fever in the probabilistic programming language Birch'', where the model describing the spread of dengue fever in ``Example: Dengue fever'' is implemented as a probabilistic program in the PPL Birch \cite{Birch2018}.
\begin{figure}[htb]
	\centering
	\begin{minipage}{.05\textwidth}
		\hspace{1pt}
	\end{minipage}%
	\begin{minipage}{.5\textwidth}
		\begin{algorithmic}[1]
			\State \texttt{y[1]=$\hspace{1mm}-$0.45;} 
			\State \texttt{y[2]=$\hspace{1mm}$3.56;} 
			\State \texttt{x[1]} $\sim$ \texttt{Gaussian(0,2);} 
			\State \texttt{observe(Gaussian(x[1],2),y[1]);} 
			\State \texttt{x[2]} $\sim$ \texttt{Gaussian(2x[1],1);} 
			\State \texttt{observe(Gaussian(x[2],2),y[2]);} 
			\State \texttt{print(x[1]);} 
		\end{algorithmic}
	\end{minipage}%
	\caption{
		A probabilistic program describing a linear-Gaussian state-space model with two observations, $\obs[1]$ and $\obs[2]$, and two states, $\state[1]$ and $\state[2]$.
		This program incorporates two types of checkpoints.
		A \emph{sample} checkpoint is implemented using the symbol $\sim$, followed by the distribution. An \emph{observe} checkpoint is denoted \texttt{observe(distribution,value)}.
	}
	\label{fig:PPlgModel}
\end{figure}

\subsection{Inference in a probabilistic program} \label{sec:InferencePPL}
When a probabilistic program is executed, it produces a set of random variables.
Each time the program is run, it might encounter a different set of variables due to the randomness inherent in the program from entering different stochastic branches.
For instance, the probabilistic program in \cref{fig:PPstochB} might sample $x=1$ the first time it is run, and $y$ is consequently sampled from a Gaussian distribution.
During the second run, it might sample $x=4$, and $y$ is in this case sampled from a Student's t distribution.
The randomness present in the probabilistic program makes the design of inference methods more complicated for probabilistic programs than for other types of models.
For instance, different executions of the same program may encounter the observations in different orders.
For an inference method like SMC, it is not immediately clear how to compute the weights when such a situation occurs.
Despite these difficulties, many inference algorithms have been adapted for use in probabilistic programs.
To provide an idea of how SMC can be implemented in a PPL, ``Background: The bootstrap particle filter in a probabilistic programming language'' describes  a PPL version of the bootstrap particle filter.
Some existing PPLs that implement versions of SMC are Anglican \cite{Anglican2014}, Birch \cite{Birch2018}, Figaro \cite{Figaro2016}, Gen \cite{Cusumano2019}, LibBi \cite{LibBi2013}, Pyro \cite{Pyro2018}, Turing \cite{Turing2018}, Venture \cite{Venture2014}, and WebPPL \cite{WebPPL2014}.
Examples of glspl{ppl} that are based on other inference methods\textemdash such as variational inference, Hamiltonian Monte Carlo and Gibbs sampling\textemdash are Church \cite{Church2008}, Edward \cite{Edward2016}, Infer.NET \cite{InferNET18}, JAGS \cite{JAGS2003}, Stan \cite{Stan2017}, and WinBUGS \cite{WinBUGS2000}.

An ideal PPL should implement a variety of different inference methods.
It should also support the user by choosing an appropriate inference method for the model automatically and tune it to obtain the best possible performance.
The choice of inference method can be based on the structure of the model \cite{Birch2018}.
For example, if the model is a (physically motivated) nonlinear state-space model like those in this article, the PPL might suggest to use SMC, provided that SMC has been implemented.
On the other hand, if the model is linear-Gaussian, the PPL should suggest using a Kalman filter/smoother.
The PPLs of today are not yet at this level of automation, where they can select a suitable inference algorithm among a large selection of algorithms and then tune it for an arbitrary model.
Nonetheless, many of the existing PPLs implement more than one inference algorithm. Although the user must specify which inference method to use, just being able to try several different methods without having to reimplement the model is a huge benefit.
Some PPLs of today even provide analytic improvements like marginalization for variance reduction \cite{Wigren2019,Obermeyer2019,Hoffman2018} automatically for models where such relations exists.
There are also some PPLs that can perform adaptive tuning of the inference method without any intervention from the user \cite{Hoffman2014}.

\begin{tcolorbox}[title=Example: Dengue fever in the probabilistic programming language Birch, breakable, enhanced, before upper={\parindent15pt\noindent}]
	This section introduces the reader to a probabilistic programming language (PPL) that uses sequential Monte Carlo (SMC) as the base method for inference, and illustrates how a physically motivated state-space model can be implemented as a probabilistic program in that PPL.

\subsubsection*{Birch}
Birch \cite{Birch2018,BirchWeb} is a Turing-complete, open-source PPL that compiles to C++.
The model and the inference methods are both encoded in the Birch language. However, they are implemented separate from each other, so the user need only specify the model and decide on which pre-implemented inference algorithm to use.
Birch is an object-oriented programming language, which allows for using classes to encode common model structures.
By using a specific model class, the user can let the inference engine know something about the structure of the model, so that it can choose a suitable inference method.
SMC is the main inference method in Birch.
Markov chain Monte Carlo methods are also implemented, and by combining these with SMC, particle Markov chain Monte Carlo methods are also supported \cite{Wigren2019, DelaySampling2018}.
Birch utilizes three types of checkpoints to perform inference: the \textit{assume} checkpoint, which initializes a random variable $X$ to have some distribution $p$; the \textit{observe} checkpoint, which conditions a random variable $X$ on the value of some other variable $Y$ being $y$; and the \textit{value} checkpoint, which triggers sampling of a random variable $X$ that was previously only assumed.

\subsubsection*{The dengue fever model in Birch}
The results for particle Gibbs (PG) presented in ``Example: Particle Gibbs for identification of dengue fever parameters'' for the model described in ``Example: Dengue fever'' were, in fact, obtained using probabilistic programming in Birch.
To replicate these results, the user must only implement the model as a probabilistic program written in Birch code, since the PG algorithm is already supported in Birch through implementations of Gibbs sampling and conditional SMC.
If the object-oriented nature of Birch is utilized, a probabilistic program for the dengue model can be based on Birch's special model class \textit{MarkovModel}.
The MarkovModel class requires three internal functions to be defined: \textit{parameter}, which defines the parameters of the probabilistic program; \textit{initial}, which initializes the probabilistic program; and \textit{transition}, which specifies the transitions and observations in the probabilistic program.
The program for the dengue model inherits from MarkovModel and must also implement these functions.
Here, a simplified version of only the parts of the transition function that illustrates the connection to the model in ``Example: Dengue fever'' is presented, in the interest of space and clarity.
The complete code for the Birch implementation of the dengue model is available in the Github repository \textit{VectorBorneDisease} \cite{GitVBD2020}.

The probabilistic program is implemented in three levels.
At the lowest level is the implementation of a basic susceptible-exposed-infectious-recovered (SEIR) model, shown in Birch code \ref{alg:BirchSEIR}, which can be used to describe the transitions for both humans and mosquitoes. 
The symbol $<\sim$ indicates simulation from the distribution to the right of the symbol, implying that this is a \textit{value} checkpoint in the program.
The symbol $<\hspace{-1mm}-$ indicates the standard assignment of a value to a variable.
The first three rows in the program samples the number of newly exposed, infectious, and recovered, respectively, according to \eqref{eq:DengueStateUpdate}.
Similarly, row 4-7 update the number of susceptible, exposed, infectious, and recovered according to \eqref{eq:DengueStateUpdate}. 
\setcounter{algorithm}{0}
\setlength{\intextsep}{0pt}
\begin{algbackbox}
	\begin{algorithm}[H]
		\floatname{algorithm}{Birch code}
		\caption{SEIRModel}
		\label{alg:BirchSEIR}
\begin{lstlisting}[mathescape,basicstyle=\scriptsize\ttfamily,breaklines=true]
final class SEIRModel < MarkovModel<SEIRParameter,SEIRState> {	
   function transition(x':SEIRState,x:SEIRState,$\theta$:SEIRParameter ,ne:Integer,ni:Integer,nr:Integer) {
      x'.$\Delta$e <$\sim$ Binomial(ne, $\theta$.$\lambda$);
      x'.$\Delta$i <$\sim$ Binomial(ni, $\theta$.$\delta$);
      x'.$\Delta$r <$\sim$ Binomial(nr, $\theta$.$\gamma$);
		
      x'.s <- x.s - x'.$\Delta$e;
      x'.e <- x.e + x'.$\Delta$e - x'.$\Delta$i;
      x'.i <- x.i + x'.$\Delta$i - x'.$\Delta$r;
      x'.r <- x.r + x'.$\Delta$r;
		
      x'.n <- x'.s + x'.e + x'.i + x'.r;
   }
}
\end{lstlisting}
	\end{algorithm}
\end{algbackbox}
\setlength{\intextsep}{12pt plus 2pt minus 2pt}	

The next level implements the interaction between the human and mosquito SEIR models, see Birch code \ref{alg:BirchVBD}.
It first defines one SEIR model (Birch code \ref{alg:BirchSEIR}) for humans and mosquitoes, respectively.
It then samples the number of susceptible humans and mosquitoes that have been exposed to the virus (bitten) according to \cref{eq:DengueTao}.
The last two rows call the transition function for each of the SEIR models as implemented in Birch code \ref{alg:BirchSEIR}.
\setlength{\intextsep}{0pt}
\begin{algbackbox}
	\begin{algorithm}[H]
	\floatname{algorithm}{Birch code}
	\caption{VBDModel}
	\label{alg:BirchVBD}
\begin{lstlisting}[mathescape,basicstyle=\scriptsize\ttfamily,breaklines=true]
final class VBDModel < MarkovModel<VBDParameter,VBDState> {
   h:SEIRModel;
   m:SEIRModel;
	
   function transition(x':VBDState,x:VBDState,$\theta$:VBDParameter) {
      ne_h:Integer;
      ne_m:Integer;
	
      ne_h <$\sim$ Binomial(x.h.s, 1.0 - exp(-x.m.i/Real(x.h.n)));
      ne_m <$\sim$ Binomial(x.m.s, 1.0 - exp(-x.h.i/Real(x.h.n)));
	
      h.transition(x'.h, x.h, $\theta$.h, ne_h, x.h.e, x.h.i);
      m.transition(x'.m, x.m, $\theta$.m, ne_m, x.m.e, x.m.i);
   }
}
\end{lstlisting}
	\end{algorithm}
\end{algbackbox}
\setlength{\intextsep}{12pt plus 2pt minus 2pt}	

The highest level, shown in Birch code \ref{alg:BirchYap}, implements the disease- and outbreak-specific parameters (not shown here) and the observation model.
First, the coupled SEIR models are defined, and the transition function in Birch code \ref{alg:BirchVBD} is called.
Then, the total number of newly infectious humans since the last observation is computed and stored in the variable $x'.z$.
Finally, if there is an observation $x'.y$, the number of newly infectious humans is conditioned on that observation according to \cref{eq:dengueObs}.
The \textit{observe} checkpoint is indicated with the symbol $\sim >$.
Note that there is no example of an \textit{assume} checkpoint, indicated by the symbol $\sim$, in Birch code \ref{alg:BirchSEIR}, \ref{alg:BirchVBD} and \ref{alg:BirchYap}.
The \textit{assume} checkpoint is typically used to initialize the parameters to have a particular distribution.
For an example, see the complete code at \cite{GitVBD2020}.

\setlength{\intextsep}{0pt}
\begin{algbackbox}
\begin{algorithm}[H] 
	\floatname{algorithm}{Birch code}
	\caption{YapDengueModel}
	\label{alg:BirchYap}
	\begin{lstlisting}[mathescape,basicstyle=\scriptsize\ttfamily,breaklines=true]
final class YapDengueModel < MarkovModel<YapDengueParameter,YapDengueState> {
   v:VBDModel;

   function transition(x':YapDengueState,x:YapDengueState ,$\theta$:YapDengueParameter) {
      v.transition(x', x, $\theta$);
      x'.z <- x.z + x'.h.$\Delta$i;
      if x'.y? {
         x'.y! $\sim$> Binomial(x'.z, $\theta$.$\rho$);
         x'.z <- 0;
      } 
   } 
}
	\end{lstlisting}
\end{algorithm}
\end{algbackbox}
\setlength{\intextsep}{12pt plus 2pt minus 2pt}	
\end{tcolorbox}

\begin{tcolorbox}[title=Background: The bootstrap particle filter in a probabilistic programming language, breakable, enhanced, before upper={\parindent15pt\noindent}]
	Here, a high-level description of how inference using a bootstrap particle filter can work in a probabilistic programming language (PPL) is presented.
This description highlights the actions that must be performed by the inference engine. However, it does not go into details on how these are implemented, since that is language-specific and beyond the scope of this article.
The interested reader is referred to the documentation for one of the PPL implementing sequential Monte Carlo (SMC) mentioned in \cref{sec:InferencePPL}.

Recall from ``Bootstrap filter'' that the bootstrap particle filter is a special case of SMC, where the state transition probabilities $p(\vstate_{\tv} \mid \vstate_{\tv-1})$ are used as proposal distributions.
In short, the bootstrap filter estimates the filtering distributions $p(\vstate_{\tv} \mid \vobs_{\range{1}{\tv}})$ using a set of weighted particles that are updated at each time step according to step (a)-(c) in \cref{alg:bootstrap-particle-fitler}. 
In PPLs, a particle corresponds to an execution of a probabilistic program.
The inference engine keeps track of $N$ parallel executions of the program that together represent the particle set.
Analogous to the standard bootstrap particle filter, where each particle has a weight, each execution of the probabilistic program has an associated weight that describes how likely the observations are, given the random choices made by that particular execution.

Each of the $N$ different executions of the program is executed deterministically from the first line of the program until a checkpoint is encountered.
When an execution of the program reaches a \textit{sample} checkpoint, the inference engine simply samples from the specified distribution\textemdash the proposal is just the state transition for a bootstrap particle filter.
This corresponds to step (b), \textit{propagate}, in \cref{alg:bootstrap-particle-fitler}.
The execution of the program then continues deterministically until the next checkpoint is encountered. 
At an \textit{observe} checkpoint, the execution of the program is paused while the inference engine performs the required actions.
To implement a bootstrap particle filter, the inference engine must perform two actions. First, it updates the weight of the execution by multiplying it with the likelihood of the new observation.
This corresponds to step (c), \textit{weight}, in \cref{alg:bootstrap-particle-fitler}.
Once all executions (particles) have updated their weights, resampling takes place.
This corresponds to step (a), \textit{resample}, in \cref{alg:bootstrap-particle-fitler}.
The resampling leads to termination of some executions and duplication (or cloning) of other executions.
After resampling, all weights are reset.
The surviving executions (and their potential clones) are then resumed.

To make the above description a bit more concrete, consider the probabilistic program in \cref{fig:PPlgModel}.
Inference in this program using a bootstrap particle filter proceeds in the following way.
First $N$ parallel executions are initiated, and the observations are stored as $\obs[1]$ and $\obs[2]$ (lines 1 and 2).
When an execution reaches the \textit{sample} checkpoint on line 3, a value for $\state[1]$ is drawn from the specified Gaussian distribution, and execution of the program is resumed.
Next, an \textit{observe} checkpoint is encountered on line 4.
The execution is paused and the inference engine updates the weight for the execution by computing the likelihood of the observation $\obs[1]$.
When weights have been computed for all $N$ parallel executions, resampling based on the weights is performed.
The surviving executions (and their eventual clones) are then resumed.
On line 5, the program encounters a new \textit{sample} checkpoint, which is handled in the same way as line 3.
Similarly, on line 6, a new \textit{observation} checkpoint is encountered where the same actions as for line 4 are performed before resuming the executions of the program.

At a first glance, it can seem like there is not much difference between a probabilistic program and a standard implementation of the particle filter.
However, note that the inference engine typically implements more than one inference method that can easily be applied to the same model.
Additionally, the execution of the probabilistic program is stochastic and might therefore encounter different variables and checkpoints in different executions.
In particular, this can occur when stochastic branching (see \cref{fig:PPstochB} for an example) is present in the probabilistic program.
Different executions of the program can then pause for weighting and resampling while being in different branches that may have a different number of observe checkpoints\textemdash the executions are not aligned.
This is something that cannot be handled by an implementation of the bootstrap particle filter in a standard programming language. 
It is problematic also for many PPLs, in the sense that many PPLs do not handle the alignment problem explicitly.
However, some work on how to automatically align observe checkpoints for SMC is available in \cite{Lunden2018}, and it is handled in a more manual fashion by processing a complete branch before resampling in Birch \cite{Birch2018}.
\end{tcolorbox}

\section{Future directions}
There are several interesting avenues for future work, a few of these are outlined here.
When it comes to modeling, an interesting direction is to develop constructions that allow for combining prior knowledge about the process with the highly successful, flexible, black-box, deep neural networks and Gaussian processes.
This would create a model where the existing information about the phenomenon under study is combined with the data in such a way that new knowledge can be gained.
A key question is how to strike the right balance between prior knowledge and new knowledge via the data.
Concrete examples of constructions of this kind are available for Gaussian processes \cite{JidlingWWS:2017} and neural networks \cite{HendriksJWS:2020} when it comes to the incorporation of linear operator constraints.
It remains a challenge to formulate models of this kind that are suitable for dynamical systems.

A subtle, but vitally important aspect in the modeling and identification techniques outlined in this article is that the model structure is assumed given.
The modeling step can require significant user input, and for many new applications it is often tedious to construct an appropriate structure.
It is interesting that many of the physically inspired models rely on relatively simple combinations of mathematical functions, in conjunction with a calculus for combining them to form models that describe the observed phenomena.
A provocative idea is to encode these atomic functions and the allowed calculus, so that the structure is learned from the data (which covers the parameter estimation idea as a special case).
Promising research in this direction can be found in~\cite{Brunton2016}. 

When it comes to the learning and inference algorithms, it remains a major challenge to handle high-dimensional systems using SMC.
Interesting developments in this direction include \cite{rebeschini2015,NaessethLS:2019,ChenSOL:2011}.

The validity of the SMC framework relies on an assumed forgetting in the dynamical system, which prevents errors from accumulating over time.
Therefore, these methods tend to struggle when there are strong and long-ranging temporal dependencies in the data (and the model).
Handling such cases is largely an open problem, although the use of ``twisting'' as discussed above can provide part of a solution.
See \cite{guarniero2017iterated, LindstenHV:2018} for developments in this direction.
Another possibility is to \emph{learn} an efficient SMC algorithm conditionally on all the available data using variational inference; see \cite{Naesseth18a,le2017,maddison2017}.

For modeling in general, the problem of using the wrong model class remains challenging, since the assumptions made at the start have a strong impact on the end result.
There are currently very few methods available that perform automatic validation of modeling assumptions, making this a highly relevant avenue for future research.
The work in \cite{LindholmZSS:2019}, where an SMC-based method is developed for model validation, provides a possible starting point.
The idea there is to make use of the generative capabilities of the model to gauge its ability to generate data that is similar to the observed data.

Finally, learning from very long observation sequences is computationally challenging, since most algorithms for identifying nonlinear state-space models require processing all observations between each parameter update, both for direct likelihood optimization and data augmentation methods.
Furthermore, some (but not all) SMC-based methods require the number of particles to grow with the number of time steps, making this computational issue even more pronounced.
Methods based on ``mini-batching'' has also been proposed for state-space identification \cite{aicher2019}. However, the temporal dependencies makes this more challenging than when learning from independent observations.

\newpage
\subsection*{Acknowledgments}
This research was partially supported by the \emph{Wallenberg AI, Autonomous Systems and Software Program (WASP)}, funded by Knut and Alice Wallenberg Foundation; the projects \emph{ASSEMBLE} (contract number: RIT15-0012)
and 
\emph{Probabilistic Modeling and Inference for Machine Learning} (contract number: ICA16-0015),
funded by the Swedish Foundation for Strategic Research (SSF); the projects \emph{NewLEADS - New Directions in Learning Dynamical Systems} (contract number: 621-2016-06079),
\emph{Handling Uncertainty in Machine Learning Systems} (contract number: 2020-04122),
and \emph{Learning flexible models for nonlinear dynamics} (contract number: 2017-03807), funded by the Swedish Research Council; ELLIIT; and \emph{Kjell och M{\"a}rta Beijer Foundation}.

\newpage 

\printbibliography


\end{document}